\documentclass[]{svjour3}                     
\usepackage[]{graphicx,subfig}
\usepackage{wrapfig}
\usepackage{sidecap}
\usepackage{float}
\usepackage{amsmath}
\usepackage{amssymb}
\usepackage{setspace}
\usepackage{epstopdf}
\usepackage{comment}
\usepackage{color}
\newcommand{\vc}{\mathbf}

\newtheorem{algorithm}{Algorithm}
\begin{document}

\title{Detecting invariant manifolds, attractors, and generalized KAM tori in aperiodically forced mechanical systems}

\author{Alireza Hadjighasem         \and\\
Mohammad Farazmand         \and
George Haller
}

\institute{A. Hadjighasem \and G. Haller \at
              Department of Mechanical Engineering \\
              McGill University, 817 Sherbrooke Ave. West, Montreal, Quebec H3A 2K6, Canada\\
              \email{alireza.hadjighasem@mail.mcgill.ca}\\           
              \email{george.haller@mcgill.ca} 
           \and
           M. Farazmand \and G. Haller \at 
           Institute for Mechanical Systems\\
           ETH Z\"urich, Tannenstrasse 3, 8092 Z\"urich, Switzerland\\
           \email{farazmand@imes.mavt.ethz.ch}\\
           \email{georgehaller@ethz.ch}
           \and
           M. Farazmand\\
           Department of Mathematics\\
           ETH Z\"urich, Rämistrasse 101, 8092 Z\"urich, Switzerland\\
           \email{farazmand@imes.mavt.ethz.ch}\\
}

\date{Received: date / Accepted: date}

\maketitle

\begin{abstract}
We show how the recently developed theory of geodesic transport barriers
for fluid flows can be used to uncover key invariant manifolds in externally
forced, one-degree-of-freedom mechanical systems. Specifically, invariant
sets in such systems turn out to be shadowed by least-stretching geodesics
of the Cauchy-Green strain tensor computed from the flow map of the
forced mechanical system. This approach enables the finite-time visualization
of generalized stable and unstable manifolds, attractors and generalized
KAM curves under arbitrary forcing, when Poincar\'{e} maps are not available.
We illustrate these results by detailed visualizations of the key
finite-time invariant sets of conservatively and dissipatively forced
Duffing oscillators.
\end{abstract}

\section{Introduction}
A number of numerical and analytical techniques are available to analyze externally
forced nonlinear mechanical systems. Indeed, perturbation methods,
Lyapunov exponents, Poincar\'{e} maps, phase space embeddings and other
tools have been become broadly used in mechanics \cite{Guckenheimer,Strogatz}.
Still, most of these techniques, are only applicable to nonlinear
systems subject to autonomous (time-independent), time-periodic, or
time-quasiperiodic forcing.

These recurrent types of forcing allow for the analysis
of asymptotic features based on a finite-time sample of the underlying
flow map--the mapping that takes initial conditions to their later
states. Indeed, to understand the phase space dynamics of an autonomous
system, knowing the flow map over an arbitrary short (but finite)
time interval is enough, as all trends can be reproduced by the repeated
applications of this short-time map. Similarly, the period map of
a time-periodic system (or a one-parameter family of flow maps for
a time-quasiperiodic system) renders asymptotic conclusions about
recurrent features, such as periodic and quasiperiodic orbits, their
stable and unstable manifolds, attractors, etc.

By contrast, the identification of key features
in the response of a nonlinear system under time-aperiodic forcing has remained an open problem.
Mathematically, the lack of precise temporal recurrence in such systems
prevents the use of a compact extended phase space on which the forced
system would be autonomous. This lack of compactness, in turn, renders
most techniques of nonlinear dynamics inapplicable. Even more importantly,
a finite-time understanding of the flow map can no longer be used
to gain a full understanding of a (potentially ever-changing) non-autonomous
system.

Why would one want to develop an understanding of mechanical systems
under aperiodic, finite-time forcing conditions? The most important
reason is that most realistic forms of forcing will take time to build
up, and hence will be transient in nature, at least initially. Even
if the forcing is time-independent, the finite-time transient
response of a mechanical system is often crucial to its design, as
the largest stresses and strains invariably occur during this period.

Similar challenges arise in fluid dynamics, where temporally
aperiodic unsteady flows are the rule rather than the exception. Observational
or numerical data for such fluid flows is only available for a limited
time interval, and some key features of the flow may only be present
for an even shorter time. For instance, the conditions creating a
hurricane in the atmosphere are transient, rather than periodic, in
nature, and the hurricane itself will generally only exist for less
than two weeks \cite{lcs_hurricane}. As a result, available asymptotic methods are clearly
inapplicable to its study, even though there is great interest in
uncovering its internal structure and overall dynamics.

In response to these challenges in fluid dynamics, a number of diagnostic
tools have been developed \cite{peacock10-1,Lai}. Only very recently,
however, has a rigorous mathematical theory emerged for dynamical structures
in finite-time aperiodic flow data \cite{haller12-1}. This theory
finds that finite-time invariant structures in a dynamical system
are governed by intrinsic, metric properties of the finite-time flow
map. Specifically, in two-dimensional unsteady flows, structures acting
as transport barriers can be uncovered with the help of geodesics
of the Cauchy-Green strain tensor used in continuum mechanics \cite{Truesdell}.
This approach generalizes and extends earlier work on hyperbolic Lagrangian Coherent
Structures (LCS), that are locally most repelling or attracting material
lines in the flow \cite{haller11-1,farazmand12-2,farazmand12-1,haller00-1}.

In this paper, we review the geodesic transport theory developed in \cite{haller12-1}
in the context of one-degree-of-freedom, aperiodically forced mechanical
systems. We then show how this theory uncovers key invariant sets
under both conservative and dissipative forcing in cases where classic
techniques, such as Poincar\'{e} maps, are not available. Remarkably,
these finite-time invariant sets can be explicitly identified as parametrized
curves, as opposed to plots requiring post-processing or feature extraction.

The organization of this paper is as follows. Section \S\ref{sec:prelim} is divided into two subsections: Section \S\ref{sec:gtheory} provides the necessary background for the \textit{\emph{geodesic theory of transport barriers}} developed in \cite{haller12-1}. In section \S\ref{sec:algs}, we describe a numerical implementation of this theory that detects finite-time invariant sets as transport barrier. Section \S\ref{sec:results} presents results from the application of this numerical algorithm to one degree-of-freedom mechanical systems. First, as a proof of concept, \S\ref{sec:periodic} considers conservative and dissipative time-periodic Duffing oscillators, comparing their geodesically extracted invariant sets with those obtained form Poincar\'{e} maps. Next, section \S\ref{sec:aperiodic} deals with invariant sets in aperiodically forced Duffing oscillators, for which Poincar\'{e} maps or other rigorous extraction methods are not available. We conclude the paper with a summary and outlook.

\section{Set-up}

\label{sec:prelim}
The key invariant sets of autonomous and time-periodic
dynamical systems--such as fixed points, periodic and quasiperiodic
motions, their stable and unstable manifolds, and attractors--are
typically distinguished by their asymptotic properties. 
In contrast, invariant sets in finite-time,
aperiodic dynamical systems solely distinguish themselves by their
observed impact on trajectory patterns over the finite time interval
of their definition. This observed impact is a pronounced lack of
trajectory exchange (or transport) across the invariant set, which remains coherent in time, i.e., only
undergoes minor deformation. Well-understood, classic examples of such transport
barriers include local stable manifolds of saddles, parallel shear
jets, and KAM tori of time-periodic conservative systems. Until recently,
a common dynamical feature of these barriers has not been identified,
hindering the unified detection of transport barriers in general non-autonomous
dynamical systems.

As noted recently in \cite{haller12-1}, however, a common feature
of all canonical transport barriers in two dimensions is that they
stretch less under the flow than neighboring curves of initial conditions
do. This observation leads to a nonstandard calculus of variations
problem with unknown endpoints and a singular Lagrangian. Below we
recall the solution of this problem from \cite{haller12-1}, with
a notation and terminology adapted to one-degree-of-freedom
mechanical oscillators.

A one-degree-of-freedom forced nonlinear oscillator can generally
be written as a two-dimensional dynamical system 
\begin{equation}
\dot{x}=v(x,t),\quad x\in U\subset\mathbb{R}^{2},\quad t\in[t_{0},t_{1}],\label{eq:dynsys}
\end{equation}
with $U$ denoting an open set in the state space, where the vector
$x$ labels tuples of positions and velocities. The vector $v(x,t)$,
assumed twice continuously differentiable, contains the velocity and
acceleration of the system at state $x$ and at time $t$. 

Let $x(t_{1};t_{0},x_{0})$ denote the final state of system (\ref{eq:dynsys})
at time $t_{1}$, given its state $x_{0}$ at an initial time $t_{0}$.
The flow map associated with (\ref{eq:dynsys}) over this time interval
is defined as 
\begin{equation}
F_{t_{0}}^{t_{1}}:\mathbf{x}_{0}\longmapsto\mathbf{x}(t_{1};t_{0},x_{0}),\label{eq:flowmap}
\end{equation}
which maps initial states to final states at $t_{1}$. The Cauchy--Green
(CG) strain tensor associated with the flow map (\ref{eq:flowmap})
is defined as 
\begin{equation}
C_{t_{0}}^{t_{1}}(x_{0})=[DF_{t_{0}}^{t_{1}}(x_{0})]^{\top}DF_{t_{0}}^{t_{1}}(x_{0}),\label{eq:cg}
\end{equation}
where $DF_{t_{0}}^{t_{1}}$ denotes the gradient of the flow map (\ref{eq:flowmap}),
and the symbol $\top$ refers to matrix transposition. 

Note that the CG tensor is symmetric and positive definite. As a result,
it has two positive eigenvalues $0<\lambda_{1}\leq\lambda_{2}$ and
an orthonormal eigenbasis $\{\xi_{1},\xi_{2}\}$. We fix this eigenbasis
so that 
\begin{align}
 & C_{t_{0}}^{t_{1}}(x_{0})\xi_{i}(x_{0})=\lambda_{i}(x_{0})\mathbf{\xi}_{i}(x_{0}),\quad\left|\xi_{i}(x_{0})\right|=1,\quad i\in\{1,2\},\nonumber \\
 & \xi_{2}(x_{0})=\Omega\xi_{1}(x_{0}),\quad\Omega=\left(\begin{array}{cc}
0 & -1\\
1 & 0
\end{array}\right).
\end{align}
We suppress the dependence of $\lambda_{i}$ and $\xi_{i}$ on $t_{0}$
and $t_{1}$ for notational simplicity.

\subsection{Geodesic transport barriers in phase space}

\label{sec:gtheory}

\begin{figure}[t]
\centering
\begin{tabular}{ccc}
 &  & \tabularnewline
 \subfloat[A repelling hyperbolic barrier (red curve) repels nearby trajectories (gray blob) exponentially fast in time.]{\includegraphics[width=.95\textwidth]{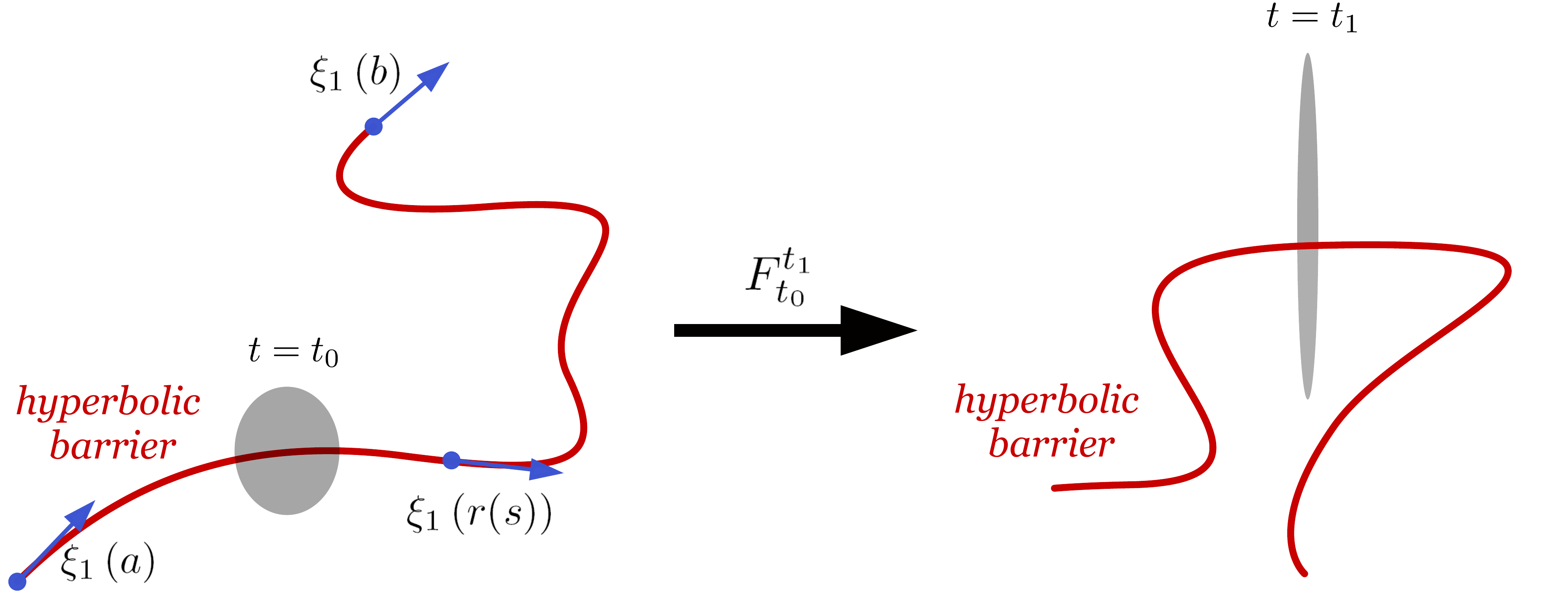}}
\tabularnewline
\subfloat[A parabolic barrier (red curve) is an open curve that has the locally largest rate of Lagrangian shear along its tangent.]{\includegraphics[width=.95\textwidth]{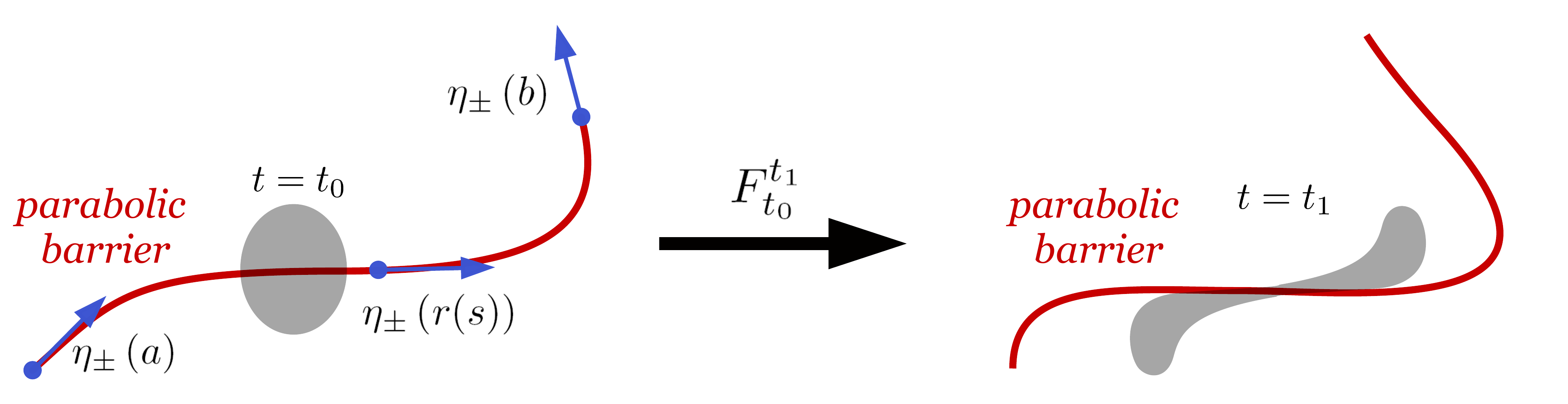}} \tabularnewline

\subfloat[An elliptic barrier (red curve) is a \emph{closed} curve with the same dynamical property as a paraolic barrier.]{\includegraphics[width=.95\textwidth]{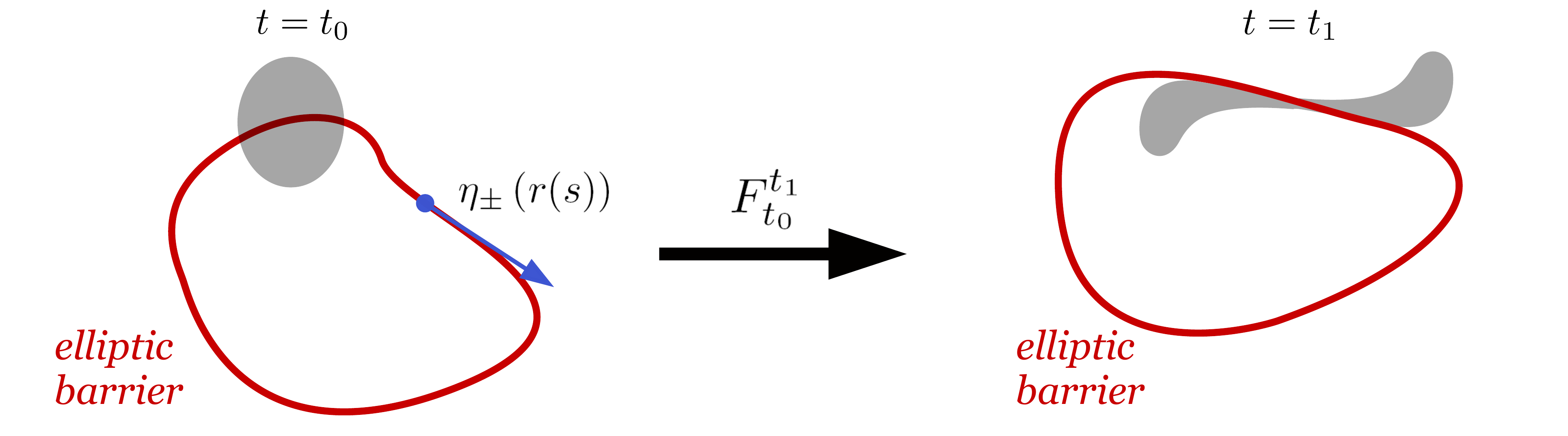}}
\end{tabular}
\caption{The three types of transport barriers in two-dimensional flows.}
\label{figure:Barriers} 
\end{figure}

A \emph{material line} $\gamma_{t}=F_{t_0}^t(\gamma_{t_0})$ is an evolving curve of initial conditions
$\gamma_{t_0}$ under the flow map $F_{t_0}^t$. As shown in \cite{haller12-1}, for
such a material line to be a locally least-stretching curve over $[t_{0},t_{1}]$,
it must be a hyperbolic, a parabolic or an elliptic line (see figure \ref{figure:Barriers}). 

The initial position $\gamma_{t_{0}}$ of a hyperbolic material line is tangent
to the vector field $\xi_{1}$ at all its points. Such material lines
are compressed by the flow by locally the largest rate, while repelling
all nearby material lines at an exponential-in-time rate. The classic
example of a hyperbolic material lines is the unstable manifold of
a saddle-type fixed point.

A parabolic material line is an open material curve whose initial position
$\gamma_{t_{0}}$ is tangent to one of the directions of locally largest
shear. At each point of the phase space, the two directions of
locally largest shear are given by 
\begin{equation}
\eta_{\pm}=\sqrt{\frac{\sqrt{\lambda_{2}}}{\sqrt{\lambda_{1}}+\sqrt{\lambda_{2}}}}\xi_{1}\pm\sqrt{\frac{\sqrt{\lambda_{1}}}{\sqrt{\lambda_{1}}+\sqrt{\lambda_{2}}}}\xi_{2},\label{eq:shrvec}
\end{equation}
as derived in \cite{haller12-1}. Parabolic material lines still repel
most nearby material lines (except for those parallel to them), but
only at a rate that is linear in time. Classic examples of parabolic
material lines in fluid mechanics are the parallel trajectories of
a steady shear flow. 

Finally, an elliptic material line is a closed curve whose initial
position $\gamma_{t_{0}}$ is tangent to one of the two directions
of locally largest shear given in (\ref{eq:shrvec}). As a result,
elliptic lines also repel nearby, nonparallel material lines at a
linear rate, but they also enclose a connected region. Classic examples of elliptic material lines are closed
trajectories of a steady, circular shear flow, such as a vortex.

Initial positions of hyperbolic material
lines are, by definition, \emph{strainlines}, i.e., trajectories of the
autonomous differential equation 
\begin{equation}
r'=\xi_{1}(r),\qquad r\in U\subset\mathbb{R}^{2},\label{eq:strline}
\end{equation}
where $r:[0,\ell]\mapsto U$ is the parametrization of the strainline
by arc-length. A \emph{hyperbolic
barrier} is then a strainline that is locally the closest to least-stretching
geodesics of the CG tensor, with the latter viewed as a metric tensor
on the domain $U$ of the phase space. The pointwise closeness of
strainlines to least-stretching geodesics can be computed in terms
of the invariants of the CG strain tensor. Specifically, the $C^{2}$
distance (difference of tangents plus difference of curvatures) of
a strainline from the least-stretching geodesic of $C_{t_{0}}^{t}$
through a point $x_{0}$ is given by the \emph{geodesic strain deviation} 

\begin{equation}
d_{g}^{\xi_{1}}(x_{0})=\frac{\left|\left\langle \nabla\lambda_{2},\xi_{2}\right\rangle +2\lambda_{2}\kappa_{1}\right|}{2\lambda_{2}^{3}},\label{eq:gdev_str}
\end{equation}
with $\kappa_{1}(x_{0})$ denoting the curvature of the strainline
through $x_{0}$ \cite{haller12-1}. A hyperbolic barrier is a compact strainline
segment on which $d_{g}^{\xi_{1}}$ is pointwise below a small threshold
value, and whose averaged $d_{g}^{\xi_{1}}$ value is locally minimal
relative to all neighboring strainlines. 

Similarly, initial positions of parabolic and elliptic material lines
are, by definition, \emph{shearlines}, i.e., trajectories of the autonomous
differential equation 
\begin{equation}
r'=\eta_{\pm}(r),\qquad r\in U\subset\mathbb{R}^{2}.\label{eq:shrline}
\end{equation}
A \emph{parabolic barrier} is an open shearline that
is close to least-stretching geodesics of the CG tensor. The pointwise
$C^{2}$-closeness of shearlines to least-stretching geodesics is
given by the \emph{geodesic shear deviation} 
\begin{eqnarray}
d_{g}^{\eta_{\pm}}(x_{0}) & = & \frac{\sqrt{1+\lambda_{2}}-\sqrt{\lambda_{1}}}{\sqrt{1+\lambda_{2}}}+\left|\frac{\left\langle \nabla\lambda_{2},\xi_{1}\right\rangle }{2\lambda_{2}\sqrt{1+\lambda_{2}}}\mp\frac{\left\langle \nabla\lambda_{2},\xi_{2}\right\rangle \left(\sqrt{1+\lambda_{2}}^{3}-\sqrt{\lambda_{2}}^{5}\right)}{2\lambda_{2}^{3}\sqrt{1+\lambda_{2}}^{3}}\right|\nonumber \\
 & \mp & \frac{\kappa_{1}\left[\sqrt{\lambda_{2}}^{5}+\left(1-\lambda_{2}^{2}\right)\sqrt{1+\lambda_{2}}\right]}{\lambda_{2}^{2}\sqrt{1+\lambda_{2}}}+\frac{\kappa_{2}}{\sqrt{1+\lambda_{2}}},\label{eq:dg_shr}
\end{eqnarray}
with $\kappa_{2}(x_{0})$ denoting the curvature of the $\xi_{2}$
vector field at the point $x_{0}$ \cite{haller12-1}. The geodesic shear deviation should
pointwise be below a small threshold level for an open shearline to
qualify as a parabolic barrier. Similarly, a closed shearline is an
\emph{elliptic barrier} if its pointwise geodesic shear deviation
is smaller than small threshold level.

For the purposes of the present discussion, we call a mechanical system of the form (\ref{eq:dynsys}) {\em conservative} if it has vanishing divergence, i.e., $\nabla\cdot v(x,t)=0$, with $\nabla$ referring to differentiation with respect to $x$. This  property implies that flow map of (\ref{eq:dynsys}) conserves phase-space area for all times \cite{arnold78}.

While a typical material line in such a conservative system will still stretch and deform significantly over time, the length of a shearline will always be preserved under the area-preserving flow map $F_{t_0}^{t_1}$ (cf. \cite{haller12-1}).  An elliptic barrier in a conservative system will, therefore, have the same enclosed area and arclength at the initial time $t_0$ and at the final time $t_1$. These two conservation properties imply that an elliptic barrier in a non-autonomous conservative system may only undergo translation, rotation and some slight deformation, but will otherwise preserve its overall shape. As a result, the interior of an elliptic barrier will not mix with the rest of the phase-space, making elliptic barriers the ideal generalized KAM curves in aperiodically forced conservative mechanical systems.

\subsection{Computation of invariant sets as transport barriers}

\label{sec:algs} In this section, we describe numerical algorithms
for the extraction of hyperbolic and elliptic barriers in a one-degree-of-freedom
mechanical system with general time dependence. Parabolic barriers
can in principle also exist in mechanical systems, but they do not
arise in the simple examples we study below. In contrast, parabolic
barriers are more common in geophysical fluid mechanics where they
typically represent unsteady shear jets.

Our numerical algorithms require a careful computation of the CG
tensor. In most mechanical systems, trajectories separate rapidly,
resulting in an exponential growth in the entries of the CG tensor.
This growth necessitates the use of a well-resolved grid, as well
as the deployment of high-end integrators in solving for the trajectories
of (\ref{eq:dynsys}) starting form this grid. Further computational
challenges arise from the handling of the unavoidable orientational
discontinuities and isolated singularities of the eigenvector fields
$\xi_{1}$ and $\xi_{2}$. The reader is referred to Farazmand \&
Haller \cite{farazmand12-1} for a detailed treatment of these computational
aspects.

As a zeroth step, we fix a sufficiently dense grid $\mathcal{G}_{0}$
of initial conditions in the phase-space $U$, then advect the grid
points from time $t_{0}$ to time $t_{1}$ under system (\ref{eq:dynsys}).
This gives a numerical representation of the flow map $F_{t_{0}}^{t_{1}}$
over the grid $\mathcal{G}_{0}$. The CG tensor field $C_{t_{0}}^{t_{1}}$
is then obtained by definition (\ref{eq:cg}) from $F_{t_{0}}^{t_{1}}$.
In computing the gradient $DF_{t_{0}}^{t_{1}}$, we use careful finite
differencing over an auxiliary grid, as described in \cite{farazmand12-1}.

Since, at each point $x_{0}\in\mathcal{G}_{0}$, the tensor $C_{t_{0}}^{t_{1}}(x_{0})$
is a two-by-two matrix, computing its eigenvalues $\{\lambda_{1},\lambda_{2}\}$
and eigenvectors $\{\xi_{1},\xi_{2}\}$ is straightforward. With the
CG eigenvalues and eigenvectors at hand, we locate the hyperbolic
barriers using the following algorithm. 
\begin{algorithm}[Locating hyperbolic barriers]
\label{alg:hyperbolic} 
\end{algorithm}

\begin{enumerate}
\item Fix a small positive parameter $\epsilon_{\xi_{1}}$ as the
admissible upper bound for the point-wise geodesic strain deviation
of hyperbolic transport barriers. 
\item Calculate strainlines by solving the ODE (\ref{eq:strline}) numerically,
with linear interpolation of the strain vector field between grid
points. Truncate strainlines to compact segments whose pointwise geodesic
strain deviation is below $\epsilon_{\xi_{1}}$ 
\item Locate hyperbolic barriers as strainline segments $\gamma_{t_{0}}$
with locally minimal relative stretching, i.e., strainline segments
that locally minimize the function 
\begin{equation}
q(\gamma_{t_{0}})=\frac{l(\gamma_{t_{1}})}{l(\gamma_{t_{0}})}.\label{eq:rel_stretch}
\end{equation}
Here $l(\gamma_{t_{0}})$ and $l(\gamma_{t_{1}})$ denote the length
of the strainline $\gamma_{t_{0}}$ and the length of its advected
image $\gamma_{t_{1}}$, respectively. 
\end{enumerate}
Computing the relative stretching (\ref{eq:rel_stretch}) of a strainline
$\gamma_{t_{0}}$, in principle, requires advecting the strainline
to time $t_{1}$. However, as
shown in \cite{haller12-1}, the length of the advected image satisfies
$l(\gamma_{t_{1}})=\int_{\gamma_{t_{0}}}\sqrt{\lambda_{1}}\, ds,$
where the integration is carried out along the strainline $\gamma_{t_{0}}$.
This renders the strainline advection unnecessary.

Numerical experiments have shown that a direct computation of $\xi_1$ is usually less accurate than that of $\xi_2$ due to the attracting nature of strongest eigenvector of the CG tensor \cite{farazmand12-1}. For this reason, computing $\xi_1$ as an orthogonal rotation of $\xi_2$ is preferable. Moreover, it has been shown \cite{faraz12_jfm} that strainlines can be computed more accurately as advected images of \emph{stretchlines}, i.e. curves that are everywhere tangent to the second eigenvector of the \emph{backward-time} CG tensor $C_{t_1}^{t_0}$. In the present paper, this approach is taken for computing the strainlines.

\begin{figure}[t]
\centering \includegraphics[width=0.7\textwidth]{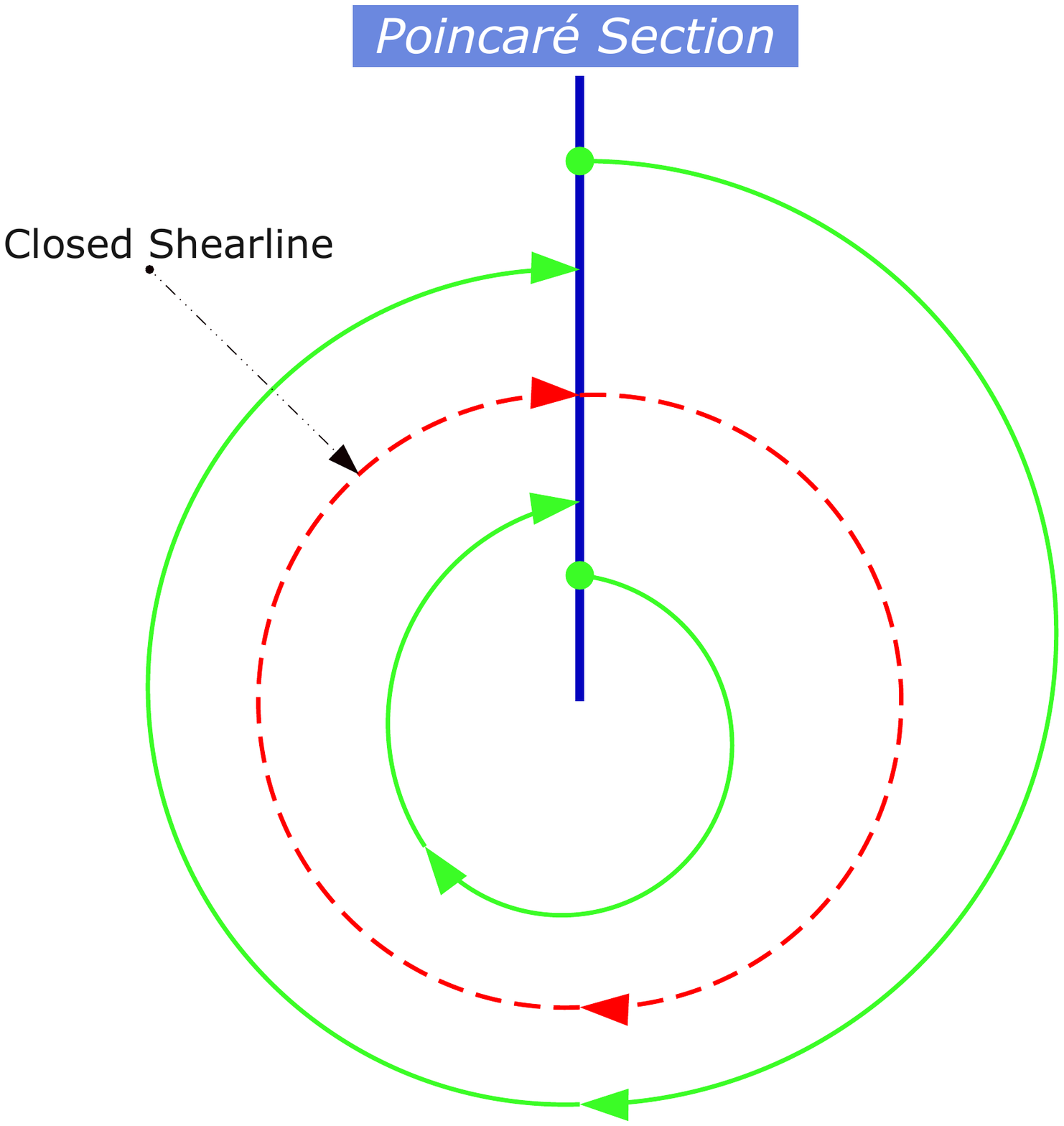}
\caption{Locating closed shearlines using a Poincar\'{e} section of the shear vector
field. Closed shearlines pass through the fixed points of the corresponding
Poincar\'{e} map.}

\label{fig:poincare} 
\end{figure}

Computing elliptic barriers amounts to finding limit cycles of the
ODE (\ref{eq:shrline}). To this end, we follow the approach used
in \cite{haller12-1,faraz12_jfm} by first identifying candidate regions for shear limit cycles
visually, then calculating the Poincar\'{e} map on a one-dimensional section
transverse to the flow within the candidate region (see figure \ref{fig:poincare}). Hyperbolic
fixed points of this map can be located by iteration, marking limit
cycles of the shear vector field (see \cite{faraz12_jfm} for more
detail).

This process is used in the following algorithm to locate elliptic
barriers. 
\begin{algorithm}[Locating elliptic barriers]
\label{alg:elliptic} 
\end{algorithm}

\begin{enumerate}
\item Fix a small positive parameter $\varepsilon_{\eta_{\pm}}$ as the
admissible upper bound for the average geodesic shear deviation of
elliptic transport barriers. 
\item Visually locate the regions where closed shearlines may exist. Construct
a sufficiently dense Poincar\'{e} map, as discussed above. Locate the
fixed points of the Poincar\'{e} map by iteration. 
\item Compute the full closed shearlines emanating from the fixed points
of the Poincar\'{e} map. 
\item Locate elliptic barriers as closed shearlines whose average geodesic
deviation $\langle d_{g}^{\eta_{\pm}}\rangle$ satisfies $\langle d_{g}^{\eta_{\pm}}\rangle<\epsilon_{\eta_{\pm}}$.
\end{enumerate}
In the next section, we use the above algorithms for locating
invariant sets in simple forced and damped nonlinear oscillators.

\section{Results}

\label{sec:results} We demonstrate the implementation of the geodesic
theory of transport barriers on four Duffing-type
oscillators. As a proof of concept, in the first two examples (section
\S\ref{sec:periodic}), we consider \emph{periodically} forced Duffing
oscillators for which we can explicitly verify our results using an appropriately
defined Poincar\'{e} map.

The next two examples deal with \emph{aperiodically} forced Duffing
oscillators (section \S\ref{sec:aperiodic}). In these examples, despite
the absence of a Poincar\'{e} map, we still obtain the key invariant sets as hyperbolic
and elliptic barriers.

To implement algorithms 1 and 2 in the forthcoming examples, the CG tensor is computed over a uniform grid $\mathcal{G}_{0}$ of $1000\times 1000$ points. A fourth order Runge-Kutta method with variable step-size (ODE45 in MATLAB) is used to solve the first-order ODEs (\ref{eq:dynsys}), (\ref{eq:strline}) and (\ref{eq:shrline}) numerically. The absolute and relative tolerances of the ODE solver are set equal to $10^{-4}$ and $10^{-6}$, respectively. Off the grid points, the strain and shear vector fields are obtained by bilinear interpolation.

In each case, the Poincar\'{e} map of algorithm \ref{alg:elliptic} is approximated by $500$ points along the Poincar\'{e} section. The zeros of the map are located by a standard secant method.

\subsection{Proof of concept: Periodically forced Duffing oscillator}

\label{sec:periodic}

\subsubsection*{Case 1: Pure periodic forcing, no damping}

Consider the periodically forced Duffing oscillator 
\begin{eqnarray*}
\dot{x}_{1} & = & x_{2},\\
\dot{x}_{2} & = & x_{1}-x_{1}^{3}+\epsilon\cos(t).
\end{eqnarray*}

For $\epsilon=0$, the system is integrable with one hyperbolic fixed
point at $(0,0)$, and two elliptic fixed points $(1,0)$ and $(-1,0)$,
respectively. As is well known, there are two homoclinic orbits connected
to the hyperbolic fixed point, each enclosing an elliptic fixed point,
which is in turn surrounded by periodic orbits. These periodic orbits
appear as closed invariant curves for the Poincar\'{e} map $P:=F_{0}^{2\pi}$.
The fixed points of the flow are also fixed points of $P$. 

\begin{figure}
\centering
\includegraphics[width=0.48\textwidth]{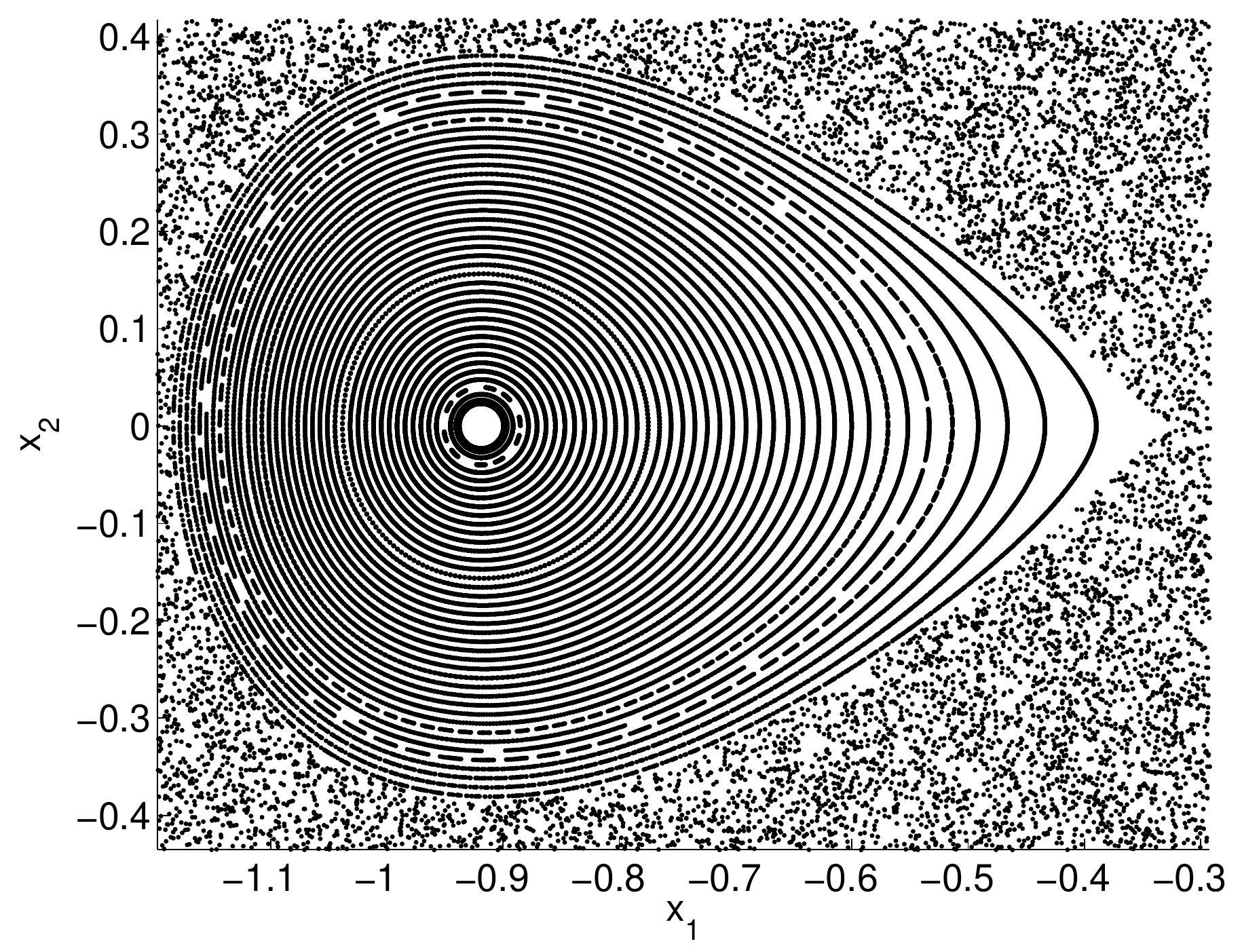} 
\includegraphics[width=0.48\textwidth]{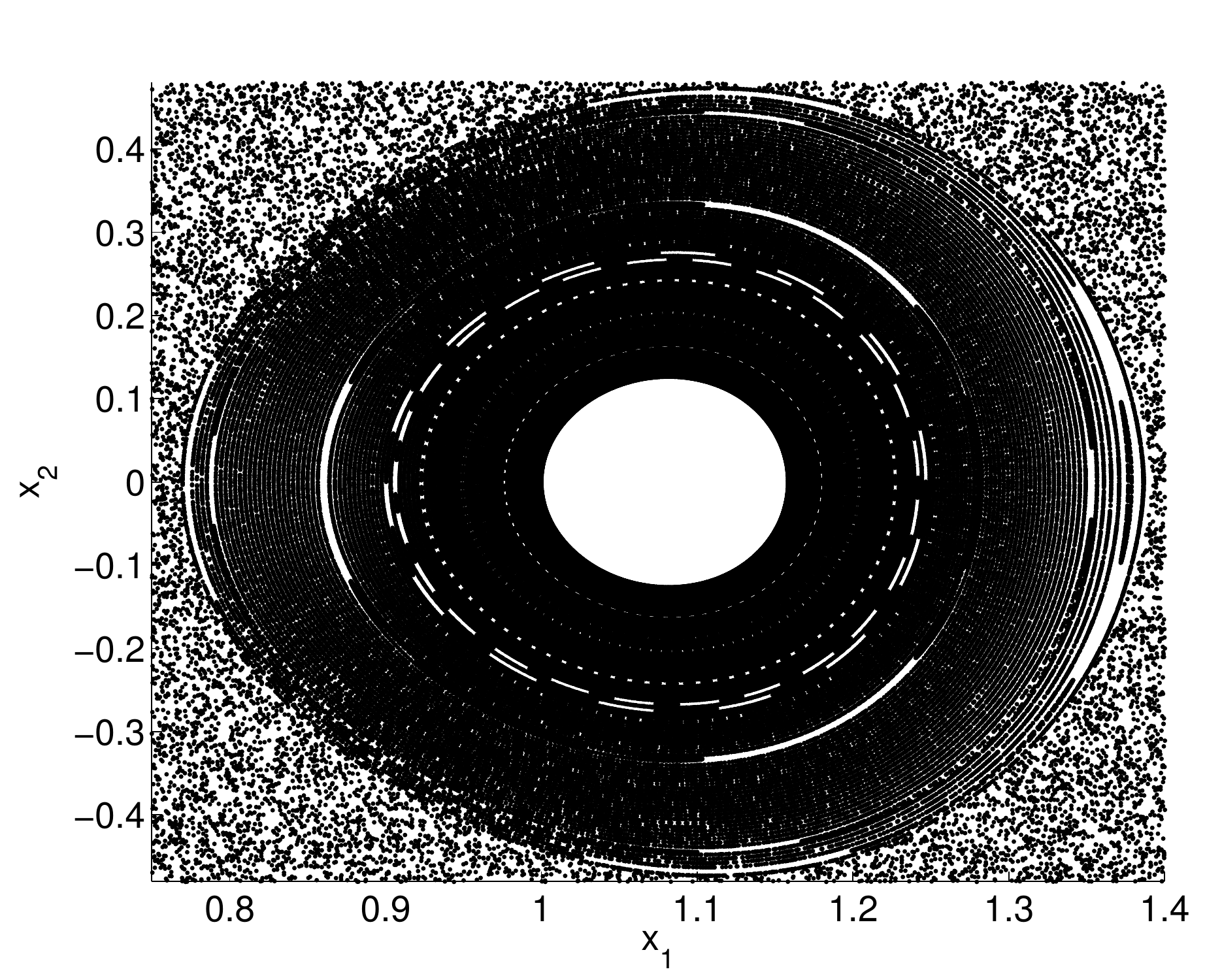}
\caption{Five hundred iterations of the Poincar\'{e} map for the periodically forced
Duffing oscillator. Two elliptic regions of the phase-space filled by
KAM tori are shown.} 
\label{figure:poincare-ex1}
\end{figure}

For $0<\epsilon\ll 1$, the Kolmogorov--Arnold--Moser (KAM) theory
\cite{arnold78} guarantees the survival of most closed invariant
sets for $P$. Figure \ref{figure:poincare-ex1} shows these surviving
invariant sets (KAM curves) of $P$ obtained for $\epsilon=0.08.$
For the KAM curves to appear continuous-looking, nearly
$500$ iterations of $P$ were needed, requiring the advection of initial
conditions up to time $t=1000\pi$. The stochastic region surrounding
the KAM curves is due to chaotic dynamics arising from the transverse
intersections of the stable and unstable manifold of the perturbed
hyperbolic fixed point of $P$.
 
The surviving KAM curves are well-known, classic examples of transport
barriers. We would like to capture as many of them as possible as
elliptic barriers using the geodesic transport theory described in
previous sections. Note that not all KAM curves are expected to prevail
as locally least-stretching curves for a given choice of the observational
time interval $[t_{0},t_{1}];$ some of these curves may take longer
to prevail due to their shape and shearing properties. 

\begin{figure}[t]
\centering
\includegraphics[width=0.48\textwidth]{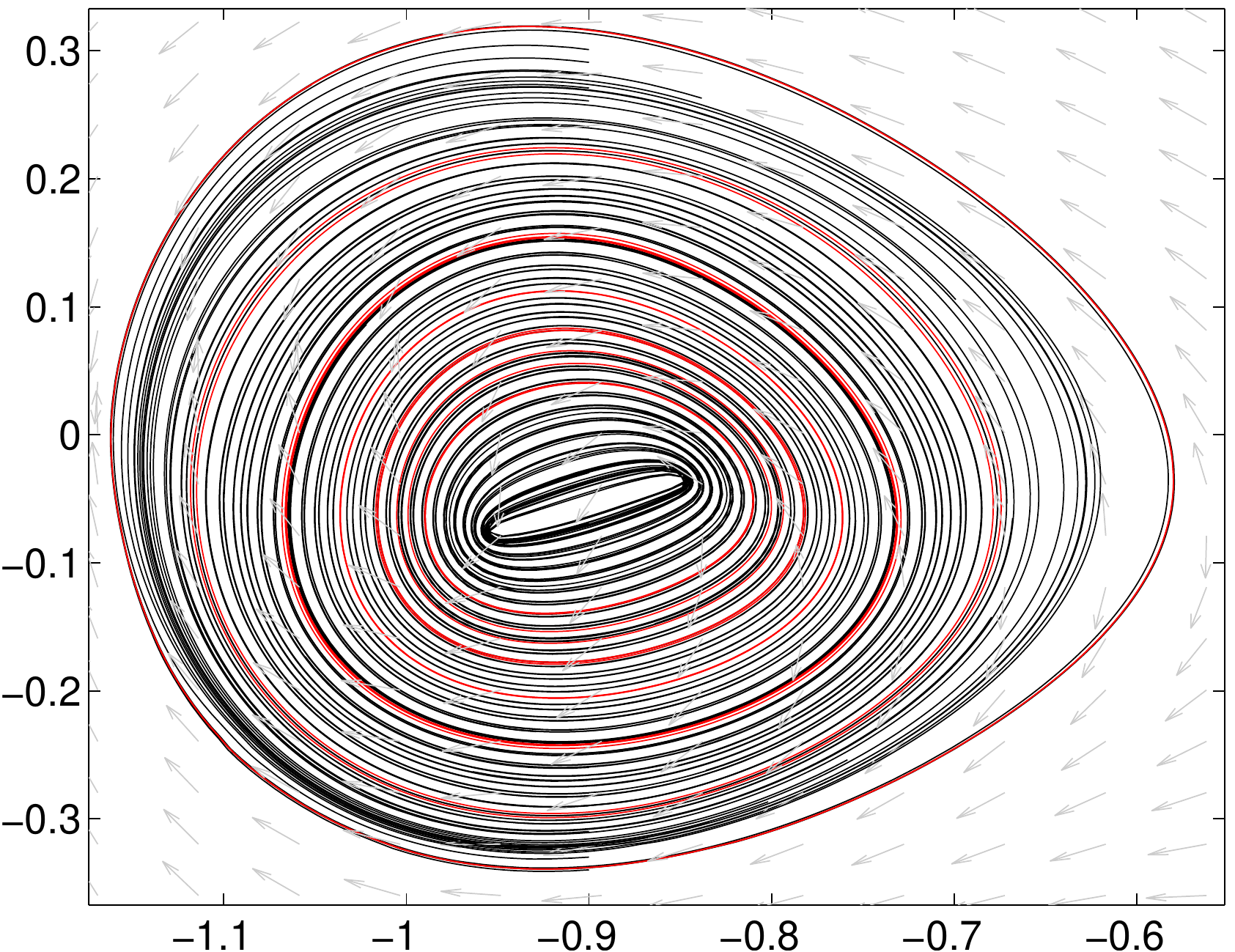} \includegraphics[width=0.48\textwidth]{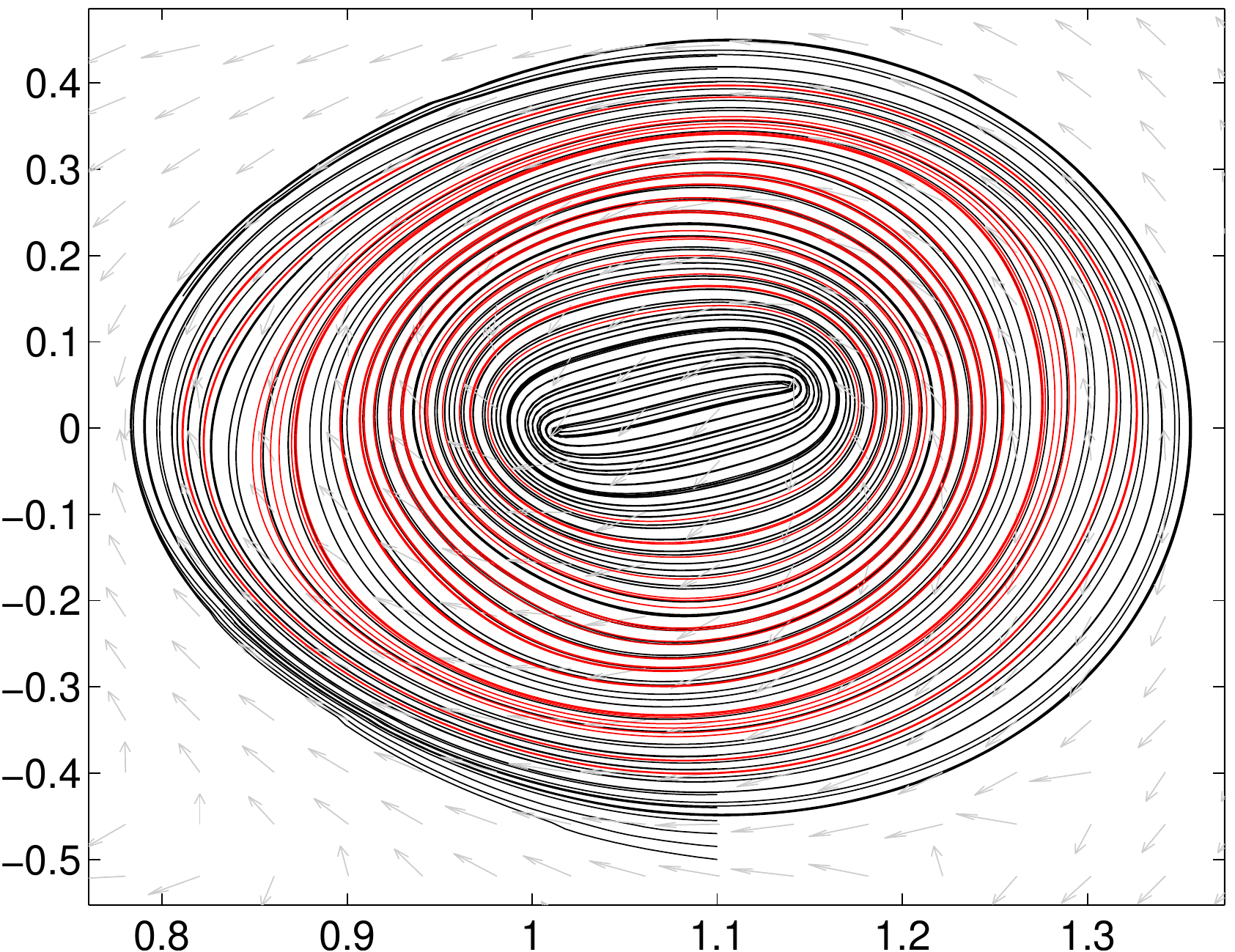}
\caption{Shearlines (black) of the periodically forced Duffing oscillator computed
at $t_{0}=0$, with integration time $T=8\pi$. The extracted elliptic
barriers with $\langle d_{g}^{\eta_{\pm}}\rangle\leq\varepsilon_{\eta_{\pm}}=0.7$ are shown in red.}
\label{figure:shearlines-ex1} 
\end{figure}

We use the elliptic barrier extraction algorithm of section \ref{sec:algs}
with $\varepsilon_{\eta_{\pm}}=0.7$. Figure \ref{figure:shearlines-ex1} shows the resulting shearlines in the KAM regions, with the closed ones marked by red. Note
that these shearlines were obtained from the CG tensor computed over
the time interval $[0,8\pi]$, spanning just four iterations of the
Poincare map. Despite this low number of iterations, the highlighted
elliptic barriers are practically indistinguishable form the KAM curves
obtained from five hundred iterations. 

\begin{figure}[t]
\centering
\begin{tabular}{ccc}
 &  & \tabularnewline
 \subfloat[$T=8\pi,$ $\langle d_{g}^{\eta_{\pm}}\rangle=0.5991.$]{\includegraphics[width=0.33\textwidth]{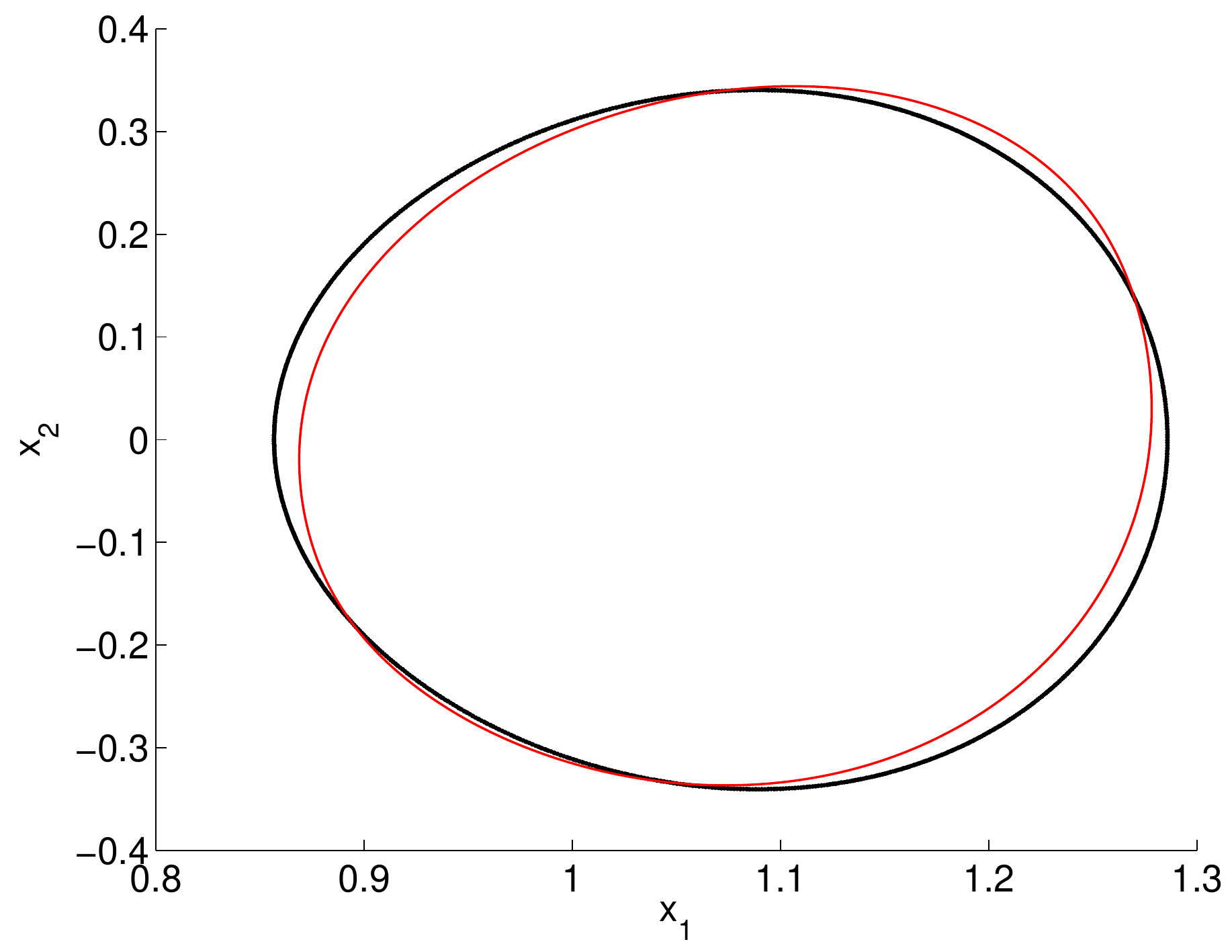}}

\subfloat[$T=10\pi,\langle d_{g}^{\eta_{\pm}}\rangle=0.3889.$]{\includegraphics[width=0.33\textwidth]{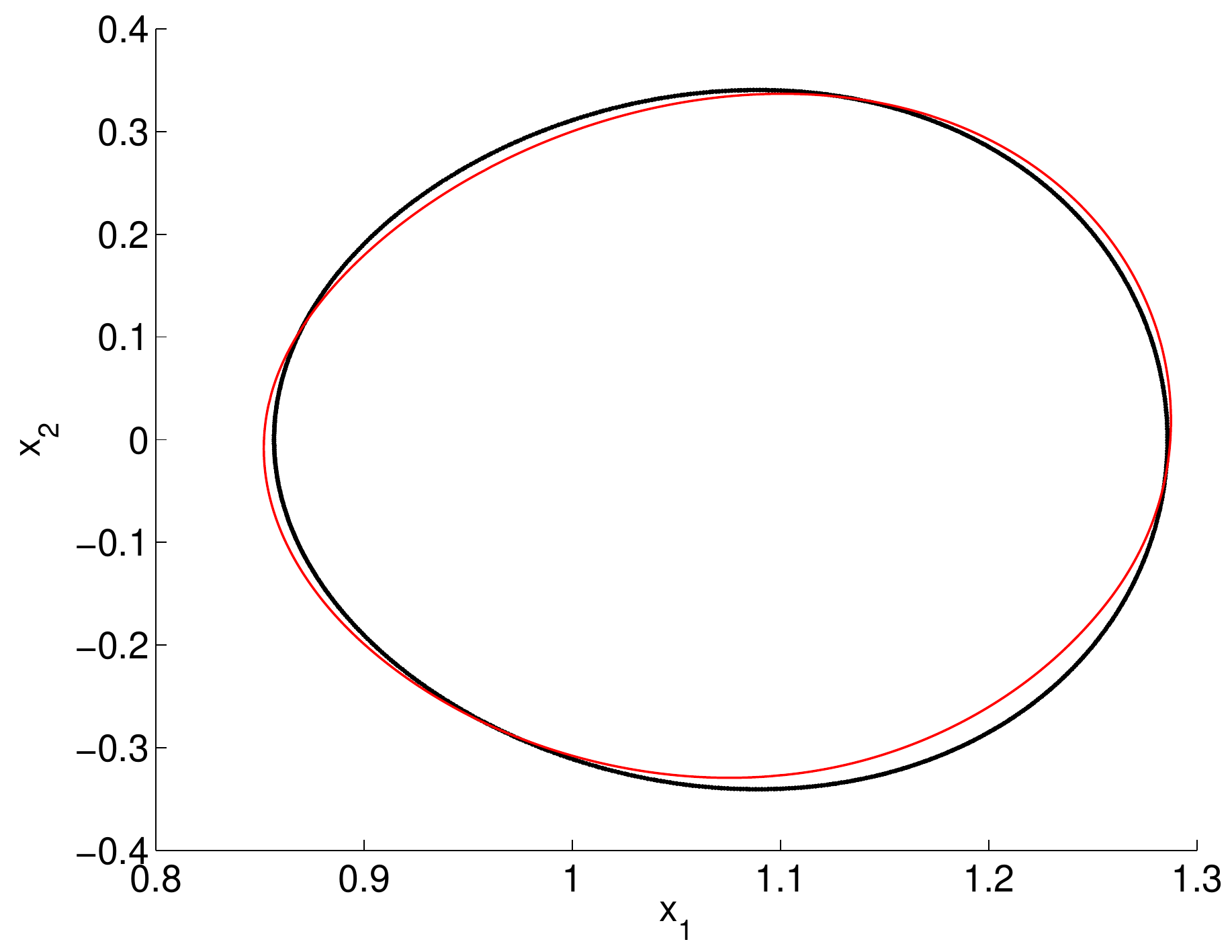}}

\subfloat[$T=12\pi$, $\langle d_{g}^{\eta_{\pm}}\rangle=0.2355.$]{\includegraphics[width=0.33\textwidth]{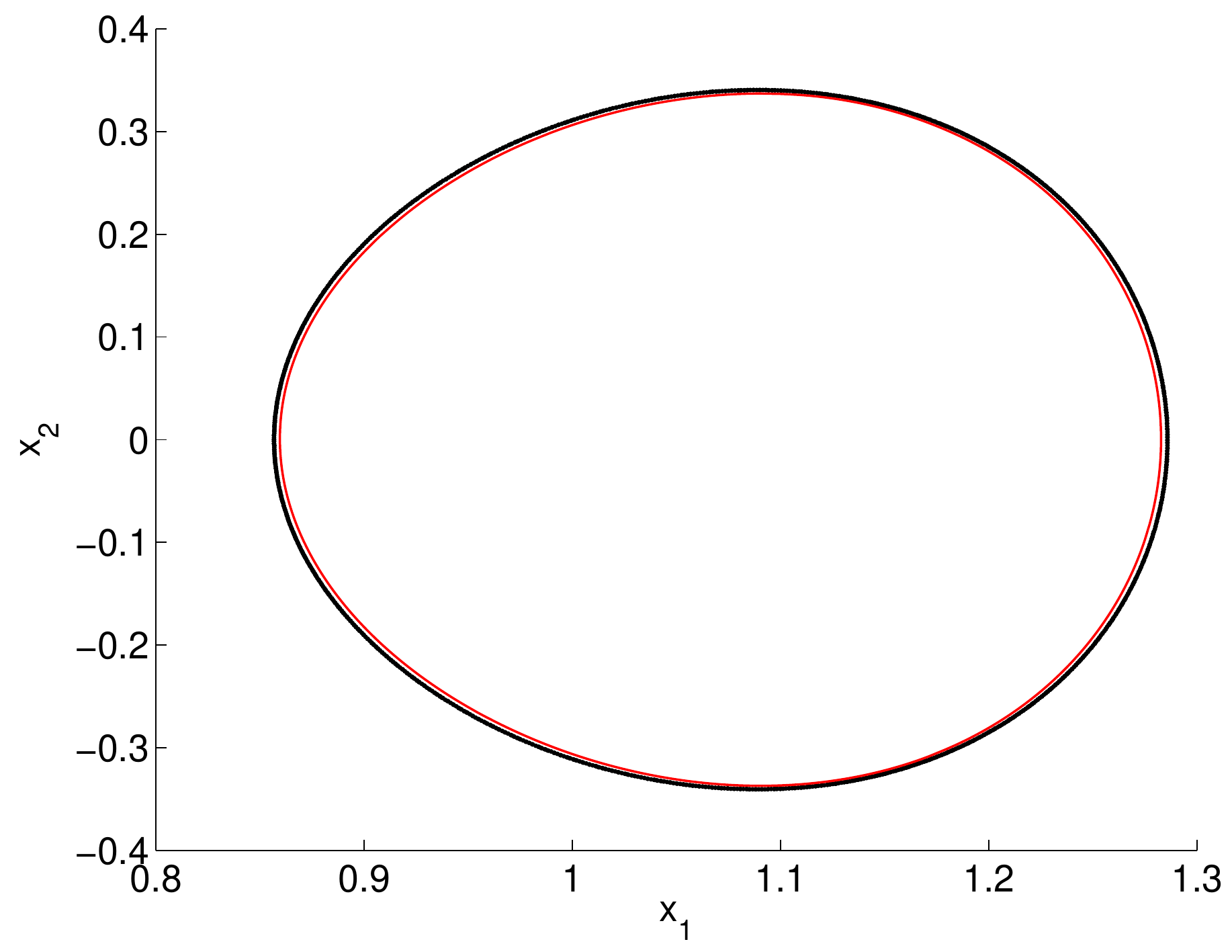}}
\end{tabular}
\caption{Convergence of an elliptic barrier (red) to a KAM curve (black) as
the integration time $T=t_{1}-t_{0}$ increases. The gradually decreasing
average geodesic deviation $\langle d_{g}^{\eta_{\pm}}\rangle$ confirms
the convergence to Cauchy--Green geodesics that closely shadow the
underlying KAM torus.}
\label{figure:poincare-shearline-ex1} 
\end{figure}

Figure~\ref{figure:poincare-shearline-ex1} shows the convergence of an elliptic
barrier to a KAM curve as the integration time $T=t_{1}-t_{0}$ increases.
Note how the average geodesic deviation $\langle d_{g}^{\eta_{\pm}}\rangle$
decreases with increasing $T$, indicating decreasing deviation from
nearby Cauchy--Green geodesics. 

Remarkably, constructing these elliptic barriers requires significantly
shorter integration time (only four forcing periods) in comparison to visualization
through the Poincar\'{e} map, which required 500 forcing periods to reveal KAM
curves as continuous objects. Clearly, the overall computational cost
for constructing elliptic barriers still comes out to be higher, since
the CG tensor needs to be constructed on a relatively dense grid $\mathcal{G}_{0}$,
as discussed in section \S\ref{sec:algs}. This high computational
cost will be justified, however, in the case of aperiodic forcing
(section \S\ref{sec:aperiodic}), where no Poincar\'{e} map is available.

\begin{figure}
\centering
\subfloat[]{\includegraphics[width=0.3\textwidth]{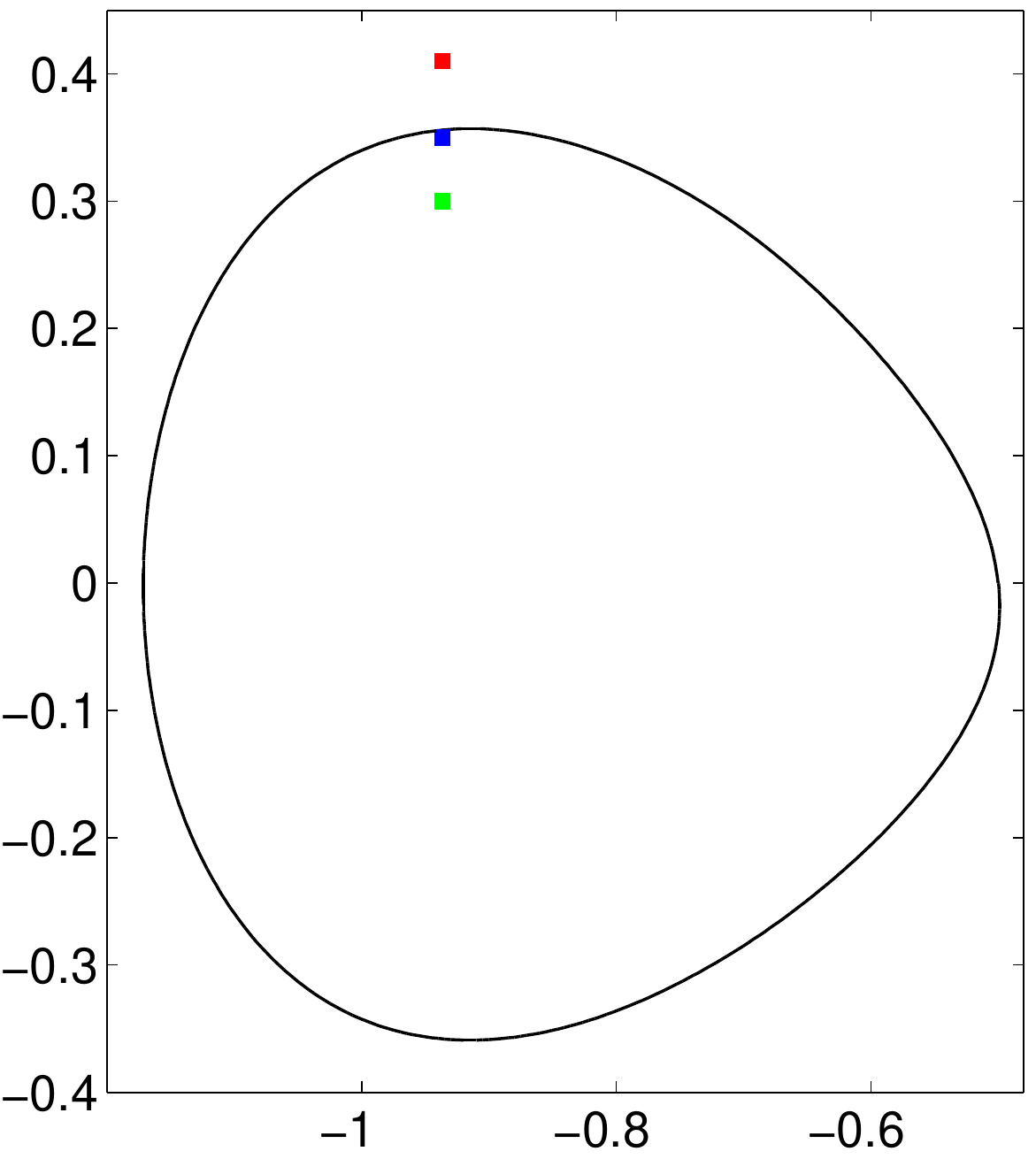}\hspace{10pt}}

\subfloat[]{\includegraphics[width=0.6\textwidth]{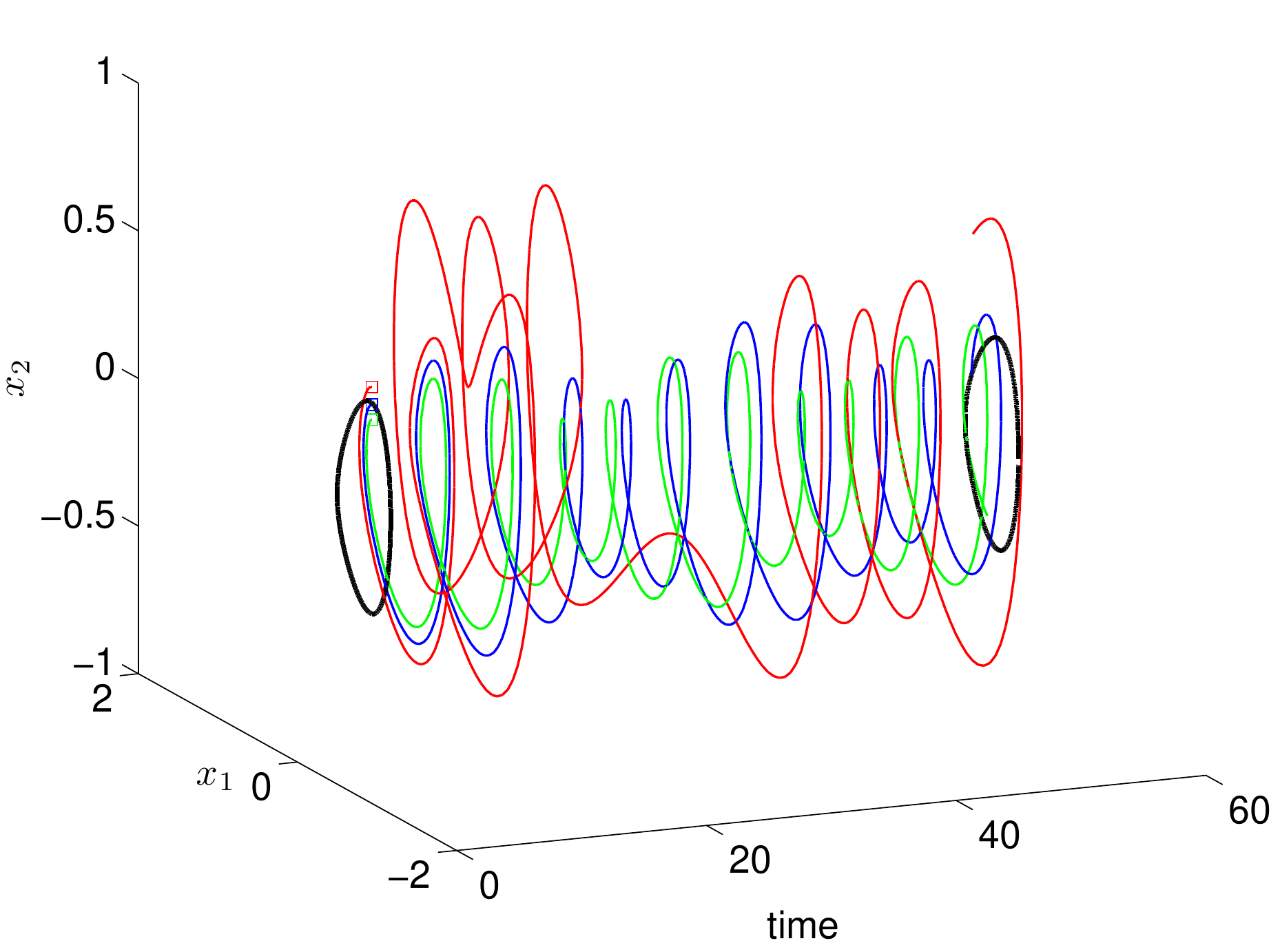}}

\caption{(a) The outermost elliptic barrier (black curve) and three initial
conditions: Two inside the elliptic barrier (blue and green) and one
outside the elliptic barrier (red). (b) The corresponding trajectories
are shown in the extended phase space of $(x_{1},x_{2},t)$. The closed
black curves mark the elliptic barrier at $t_{0}=0$ and $t_{1}=16\pi$.}
\label{figure:stability} 
\end{figure}

\begin{SCfigure} \centering \includegraphics[width=0.5\textwidth]{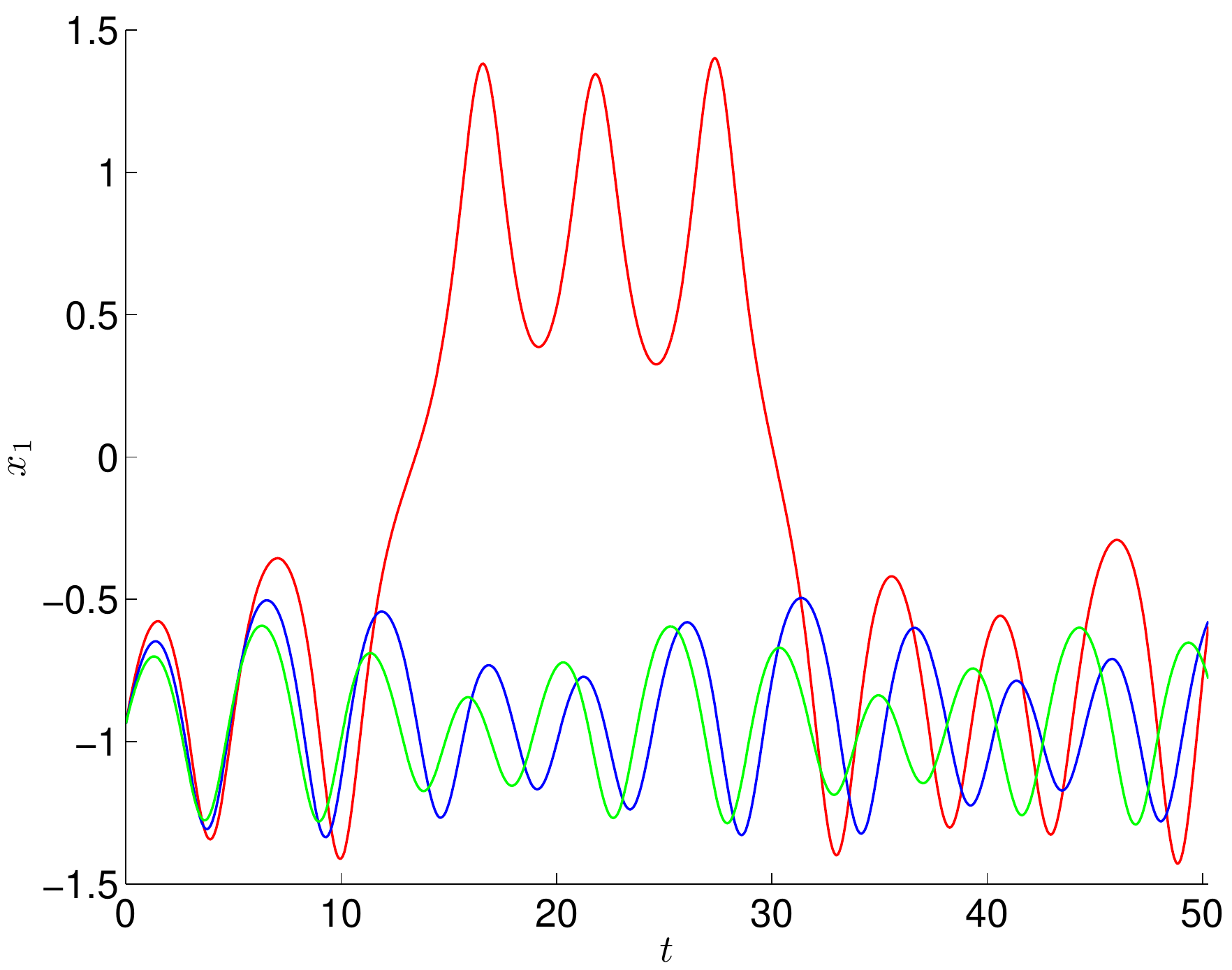}
\caption{The $x_{1}$-coordinate of the trajectories of figure \ref{figure:stability}.}
\label{figure:stability-x1} 
\end{SCfigure}

In the context of one-degree-of-freedom mechanical systems, the outermost
elliptic barrier marks the boundary between regions of chaotic dynamics
and regions of oscillations that are regular on a macroscopic scale.
To demonstrate this sharp dividing property of elliptic barriers,
we show the evolution of system (\ref{eq:duffing-ex3}) from three
initial states, two of which are inside the elliptic region and one
of which is outside (figure \ref{figure:stability}a). The system
exhibits rapid changes in its state when started from outside the
elliptic region. In contrast, more regular behavior is observed for
trajectories starting inside the elliptic region. This behavior is
further depicted in figure \ref{figure:stability-x1}, which shows
the evolution of the $x_{1}$-coordinate of the trajectories as a function of
time.

\subsubsection*{Case 2: Periodic forcing and damping}

Consider now the damped-forced Duffing oscillator
\begin{eqnarray}
\dot{x}_{1} & = & x_{2},\nonumber \\
\dot{x}_{2} & = & x_{1}-x_{1}^{3}-\delta x_{2}+\epsilon\cos(t),\label{eq:duffing_dissipative}
\end{eqnarray}
with $\delta=0.15$ and $\epsilon=0.3$. This system is known to have
a chaotic attractor that appears as an invariant set of the the Poincar\'{e}
map $P=F_{0}^{2\pi}$ (see, e.g., \cite{Guckenheimer}).
Here, we show that the attractor can be very closely approximated
by hyperbolic barriers computed via algorithm \ref{alg:hyperbolic}.

\begin{figure}
\centering
\begin{tabular}{ccc}
\subfloat[Strainlines computed for the damped-forced Duffing oscillator (\ref{eq:duffing_dissipative})
at time $t_{0}=0$, with the integration time $T=-8\pi$.]{\includegraphics[width=0.5\textwidth]{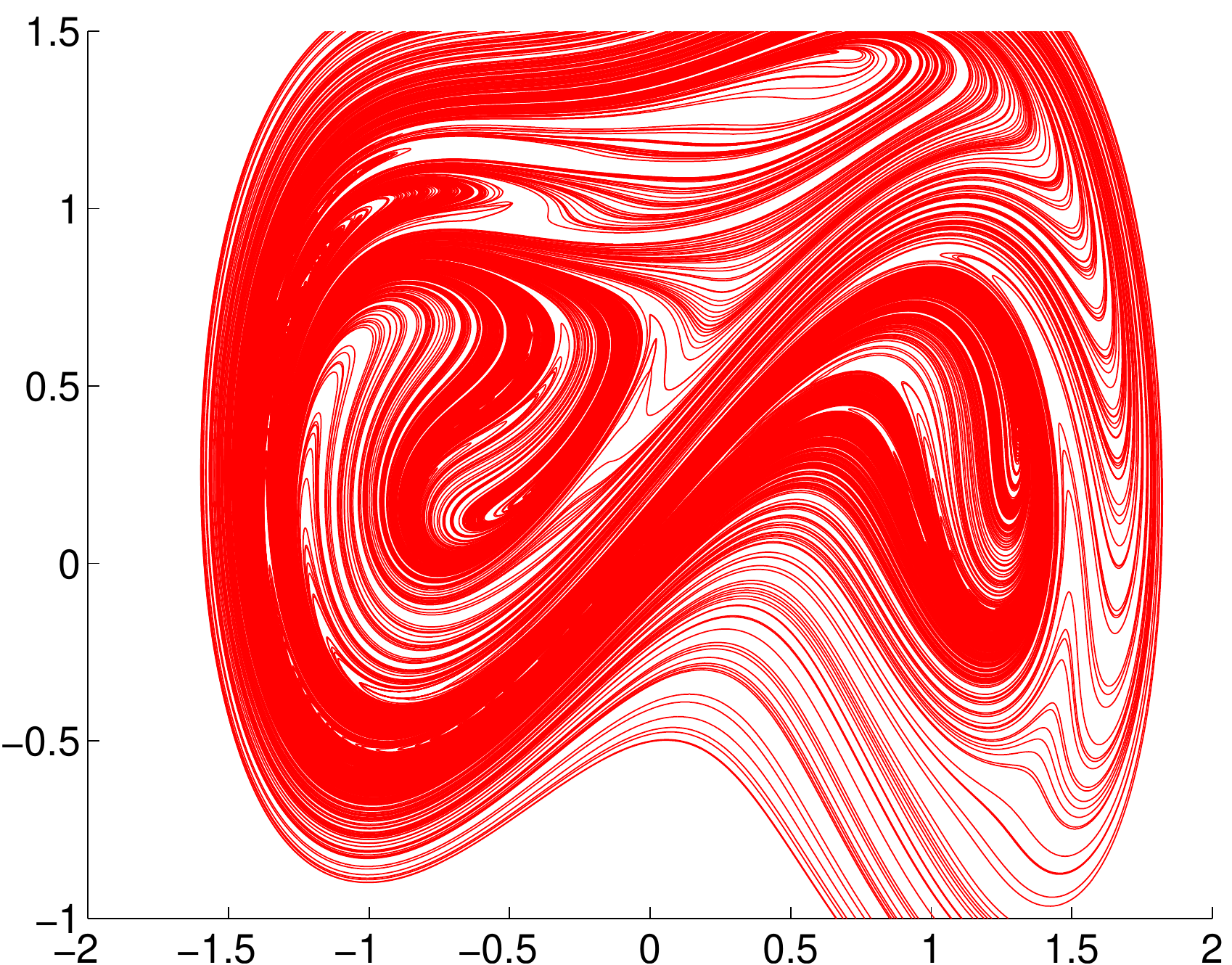}} & \tabularnewline
\subfloat[The strainline (red) with globally minimum relative stretching.
Points with $d_{g}^{\xi_{1}}>10^{-3}$ are highlighted as black dots.]{\includegraphics[width=0.5\textwidth]{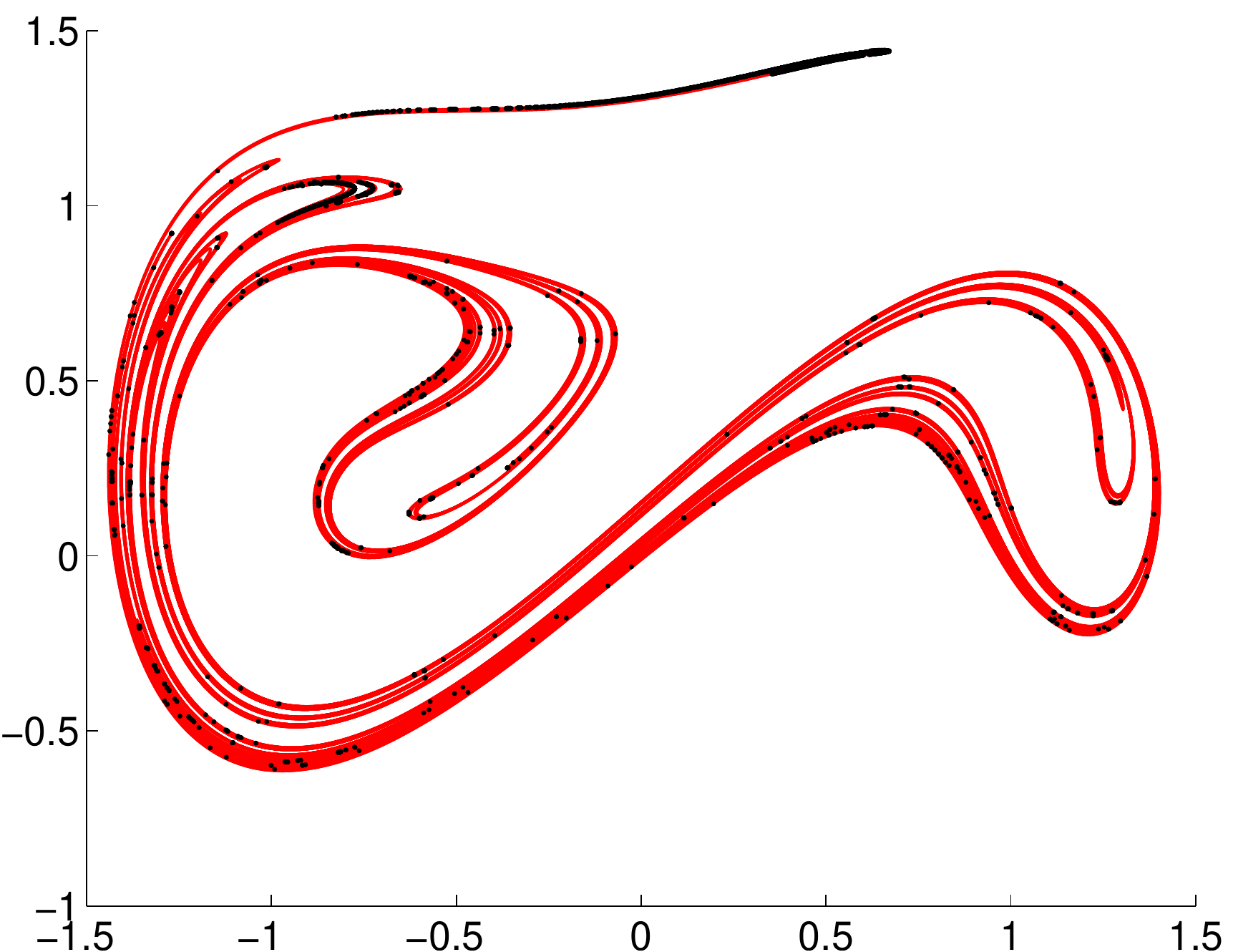}}
\hspace{12 pt}
\subfloat[Final approximation of the chaotic attractor by a single, continuous
strainline with minimal geodesic deviation.]{\includegraphics[width=0.5\textwidth]{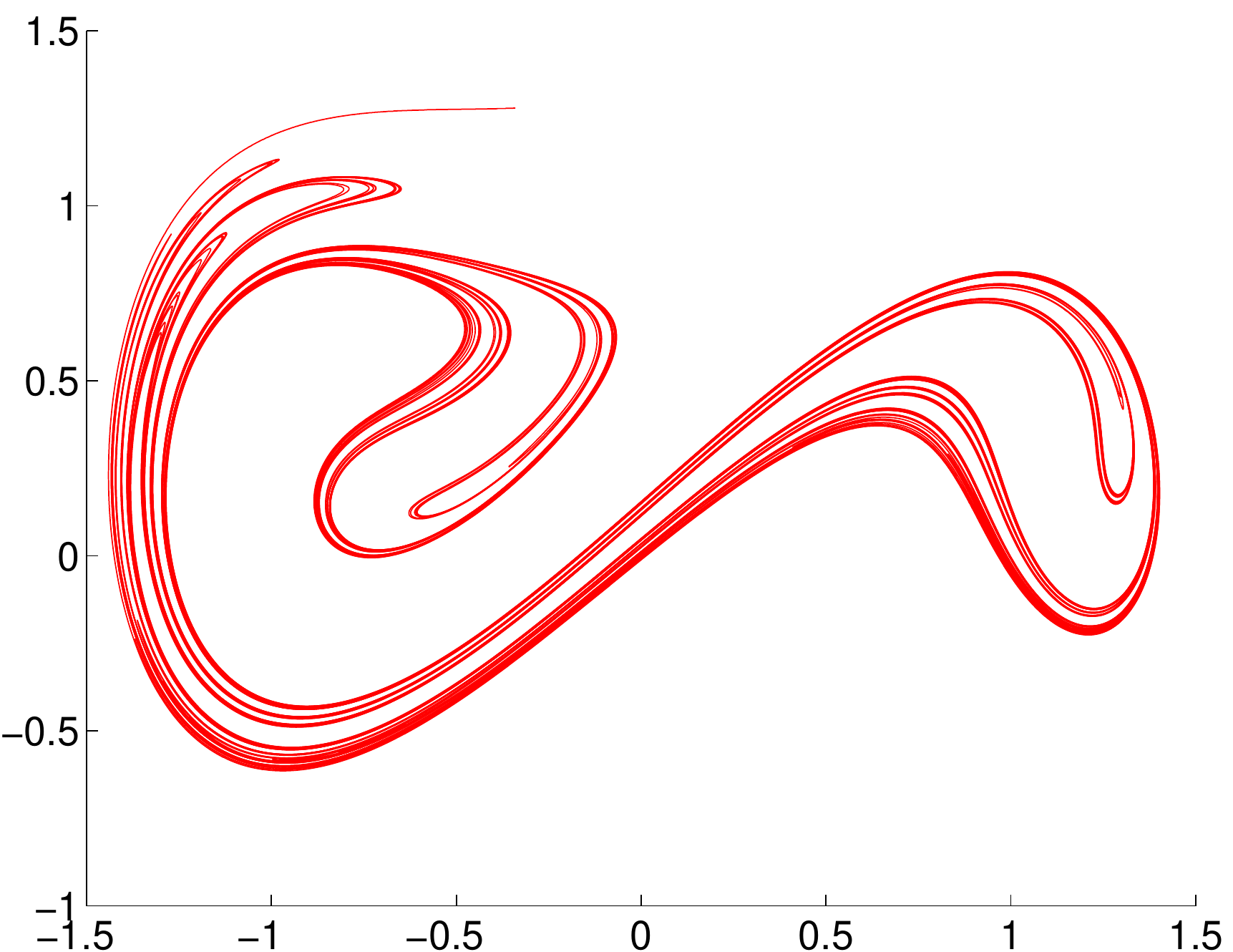}}
\end{tabular}
\caption{Construction of the attractor of the damped-forced Duffing oscillator as a hyperbolic transport barrier.}
\label{figure:strainline-ex2} 
\end{figure}

Figure~\ref{figure:strainline-ex2}a shows strainlines computed backward
in time with $t_{0}=0$ and integration time $T=t_{1}-t_{0}=-8\pi$.
The strainline with globally minimal relative stretching (\ref{eq:rel_stretch})
is shown in figure~\ref{figure:strainline-ex2}b. Black dots
mark the points where the geodesic deviation $d_{g}^{\xi_{1}}$ exceeds
the admissible upper bound $\epsilon_{\xi_{1}}=10^{-3}$. At its tail
(covered by black dots), the strainline persistently deviates
from CG geodesics, and hence should be truncated. The resulting hyperbolic barrier, as a finite-time
approximation to the chaotic attractor, is shown in figure~\ref{figure:strainline-ex2}c.

\begin{figure}
\begin{centering}
\begin{tabular}{cc}
 & \tabularnewline
\subfloat[]{\includegraphics[width=0.5\textwidth]{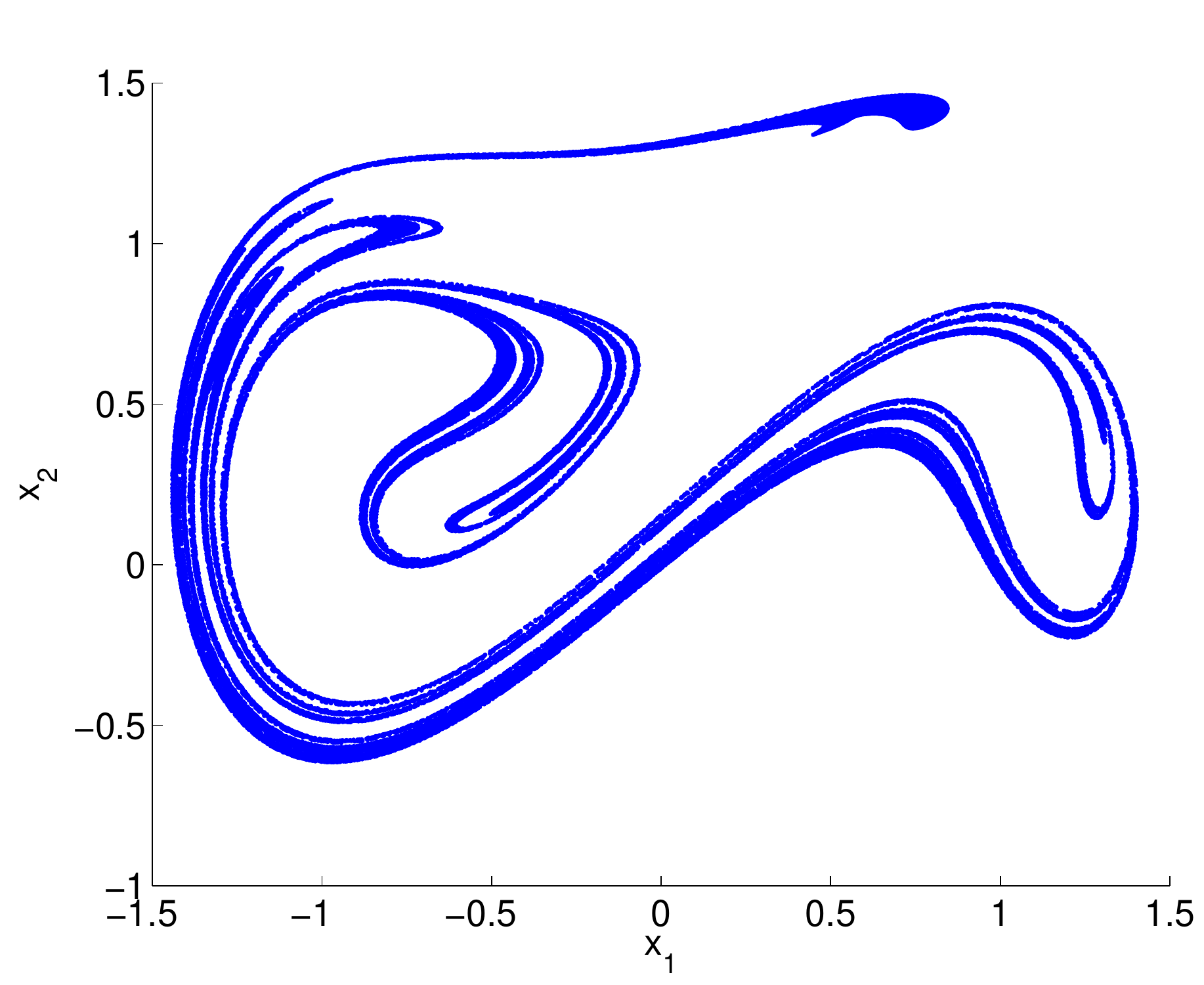}}

\subfloat[]{\includegraphics[width=0.5\textwidth]{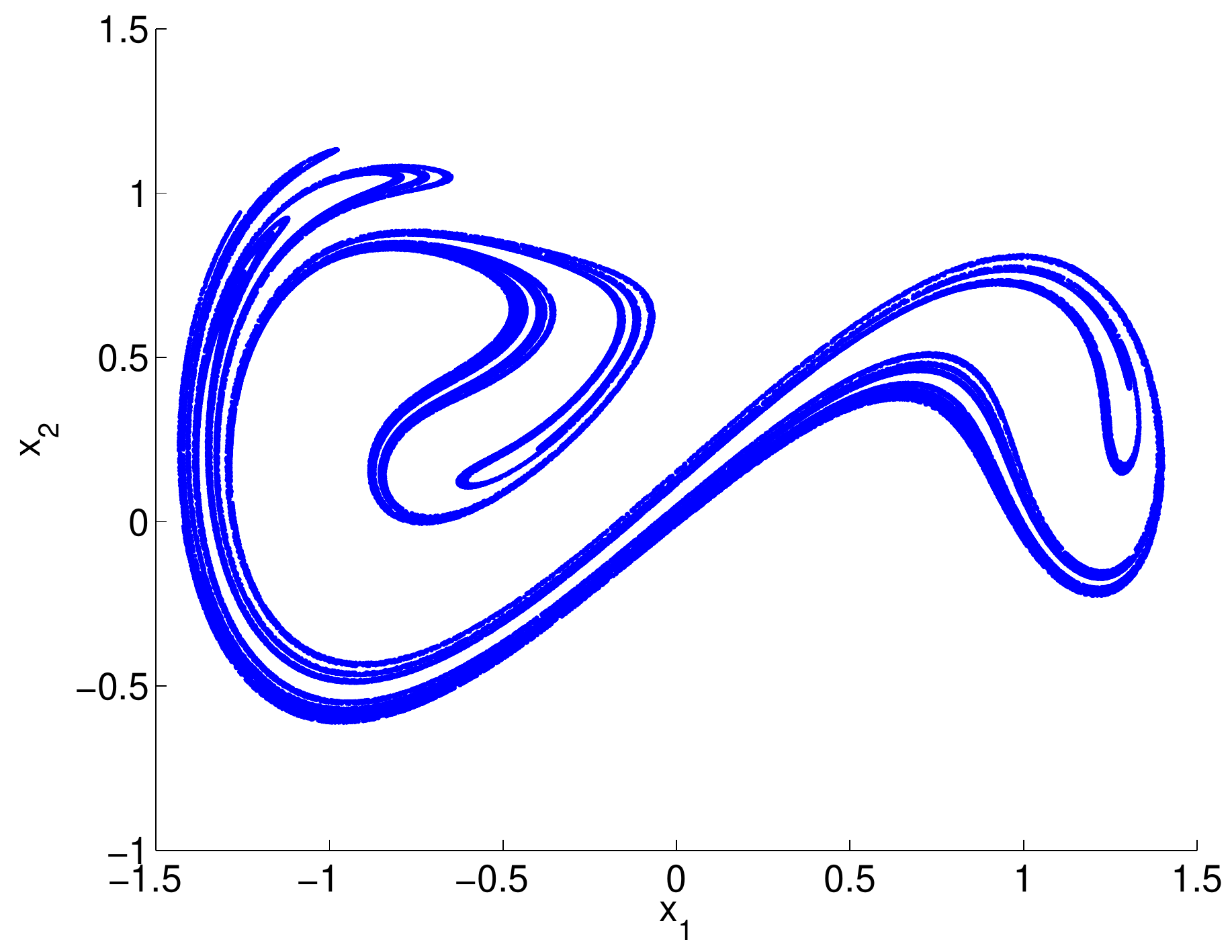}} & \tabularnewline
\subfloat[]{\includegraphics[width=0.5\textwidth]{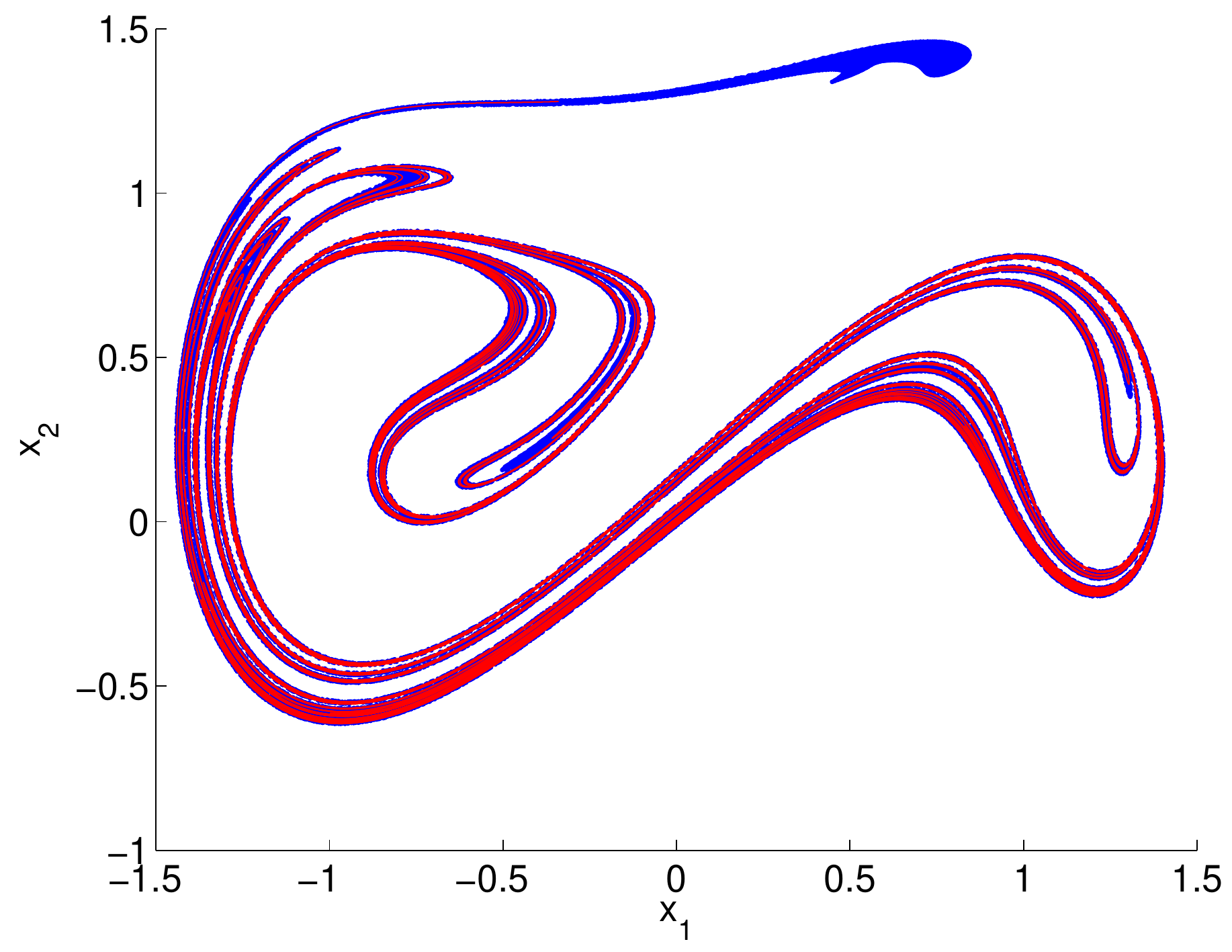}}

\subfloat[]{\includegraphics[width=0.5\textwidth]{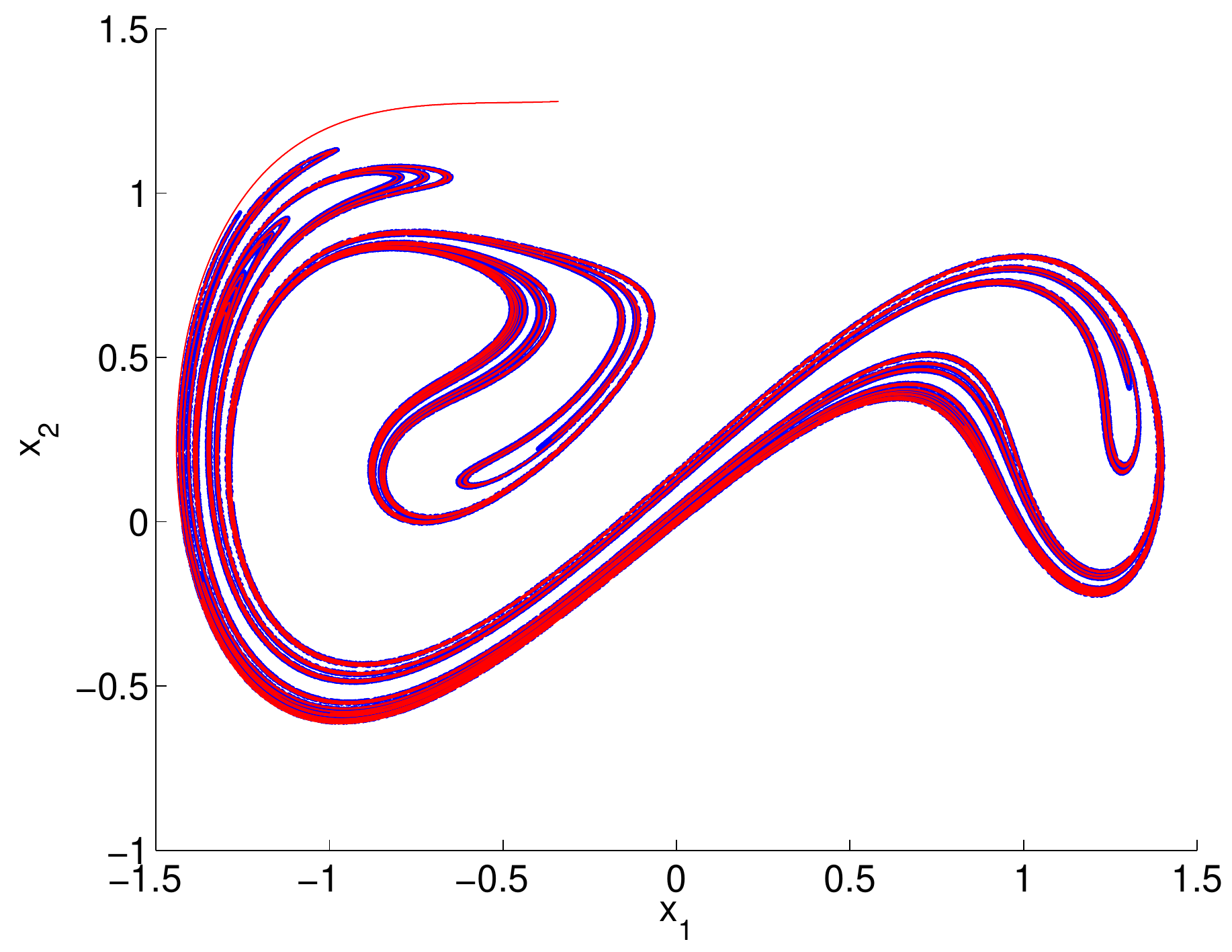}

} & \tabularnewline
\end{tabular} 
\par\end{centering}

\caption{(a) Attractor of system (\ref{eq:duffing_dissipative}) obtained from
four iterates of the Poincar\'{e} map. (b) Attractor obtained from 20 iterates
of the Poincar\'{e} map. (c) Attractor computed as a hyperbolic barrier
(red), compared with the Poincar\'{e} map (blue) computed for
the same integration time (four iterates). (d) Comparison of attractor
computed as a hyperbolic barrier (red) with the one obtained from 20 iteration of the Poincar\'{e} map (blue).
The integration time for locating the hyperbolic barrier is $T=t_1-t_0=-8\pi$.}

\label{figure:strainline-poincare-ex2} 
\end{figure}

The approximate location of the attractor can also be revealed by
applying the Poincar\'{e} map to a few initial conditions (tracers) released
from the basin of attraction. For long enough advection time, the
initial conditions converge to the attractor highlighting its position
(see figure \ref{figure:strainline-poincare-ex2}a and \ref{figure:strainline-poincare-ex2}b). In figure \ref{figure:strainline-poincare-ex2}c,
the hyperbolic barrier is superimposed on the advected tracers showing close agreement between the two. Figure \ref{figure:strainline-poincare-ex2}d shows the tracers advected for a longer time ($T=40\pi$) together
with the hyperbolic barrier; the two virtually coincide. Note that the hyperbolic barrier is a smooth, parametrized curve (computed as a trajectory of (\ref{eq:strline})), while the tracers form a set of scattered points.

\subsection{The aperiodically forced Duffing oscillator}\label{sec:aperiodic} 

In the next two examples, we study aperiodically
forced Duffing oscillators. In the presence of aperiodic forcing, the
Poincar\'{e} map $P$ is no longer defined as the system lacks any recurrent
behavior. However, KAM-type curves (i.e., closed curves, resisting significant
deformation) and generalized stable and unstable manifolds (i.e., most
repelling and attracting material lines) exist in the phase-space
and determine the overall dynamics of the system.

\subsubsection*{Case 1: Purely aperiodic forcing, no damping}

\begin{SCfigure}
\centering \includegraphics[scale=0.4]{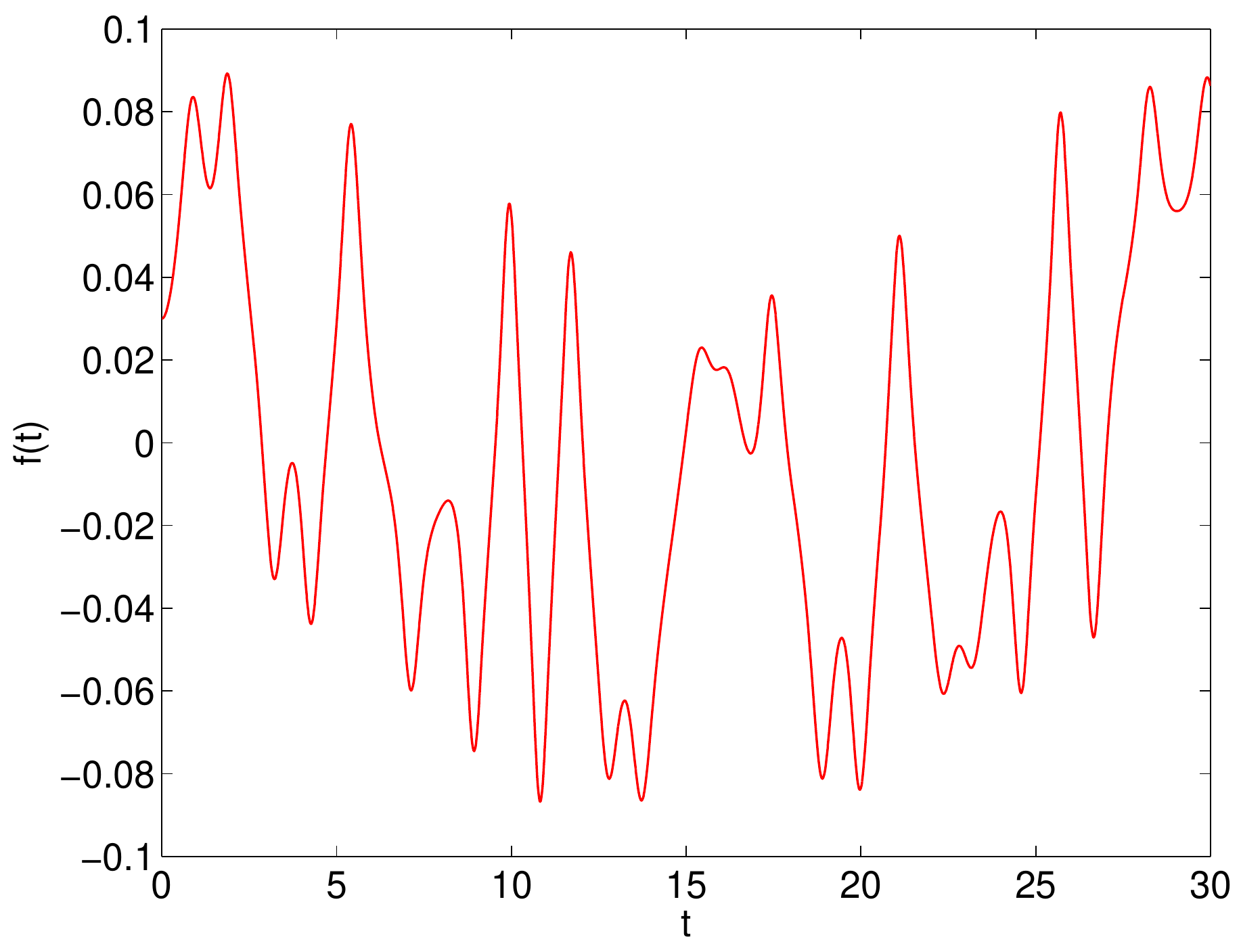} \caption{Chaotic forcing function $f(t)$ for equation (\ref{eq:duffing-ex3}).}
\label{figure:force-ex3} 
\end{SCfigure}

Consider the Duffing oscillator 
\begin{eqnarray}
\dot{x}_{1} & = & x_{2},\nonumber \\
\dot{x}_{2} & = & x_{1}-x_{1}^{3}+f(t),\label{eq:duffing-ex3}
\end{eqnarray}
where $f(t)$ is an aperiodic forcing function obtained from a chaotic
one-dimensional map (see figure \ref{figure:force-ex3}). 

While, KAM theory is no longer applicable, one may still expect KAM-type barriers
to survive for small forcing amplitudes. Such barriers would no longer be repeating themselves periodically
in the extended phase space. Instead, a generalized KAM barrier is expected to be an invariant
cylinder, with cross sections showing only minor deformation. The
existence of such structures can, however, be no longer studied via
Poincar\'{e} maps.

\begin{figure}
\centering
\subfloat[]{\includegraphics[width=0.5\textwidth]{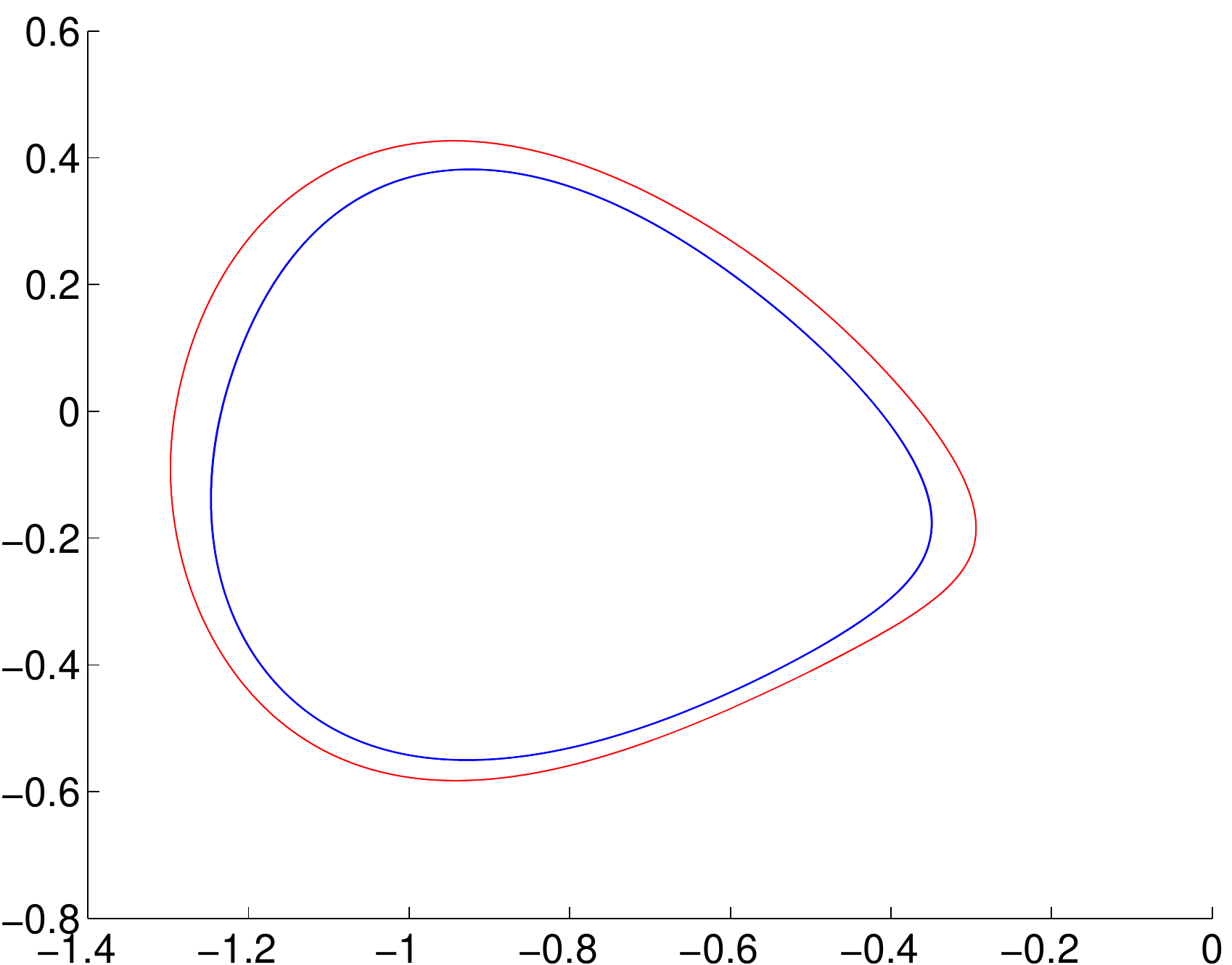} }
\subfloat[]{\includegraphics[width=0.5\textwidth]{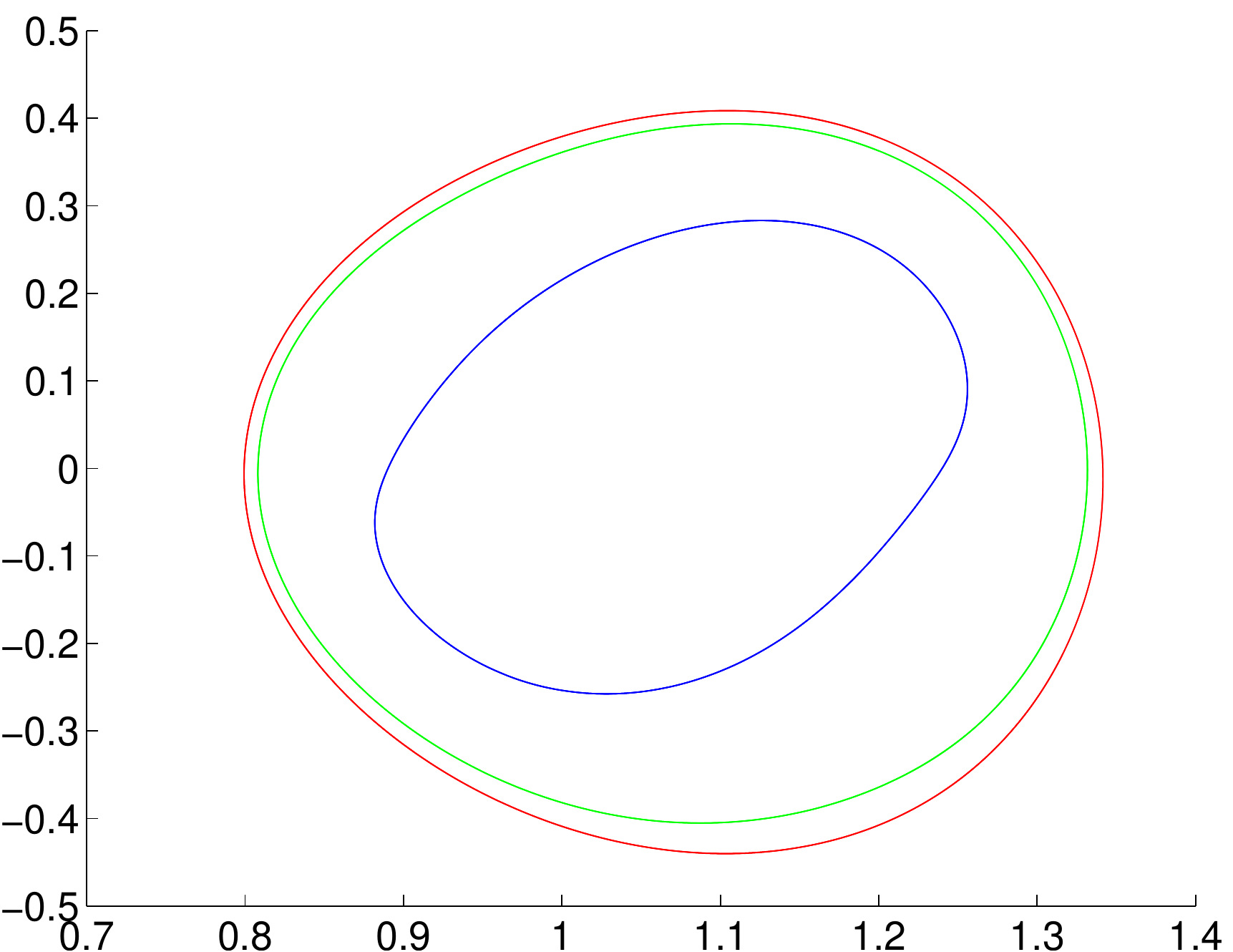}}
\caption{Closed shearlines for equation (\ref{eq:duffing-ex3}) computed in
two elliptic regions. The figure shows the shearlines at time $t_{0}=0$.
The integration time is $T=4\pi$.}
\label{figure:shearlines-ex3} 
\end{figure}

\begin{figure}
\centering
\includegraphics[scale=0.33]{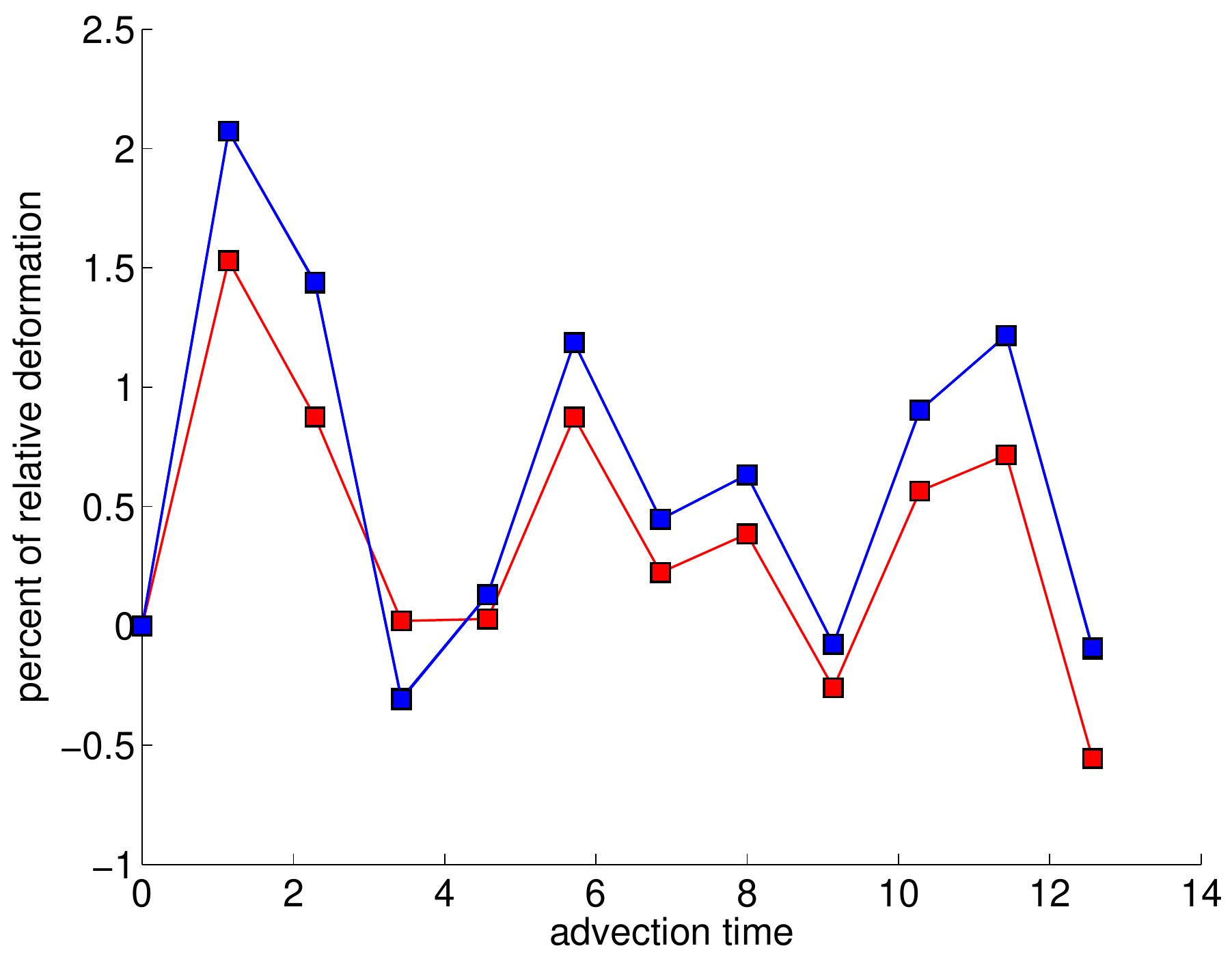} \includegraphics[scale=0.33]{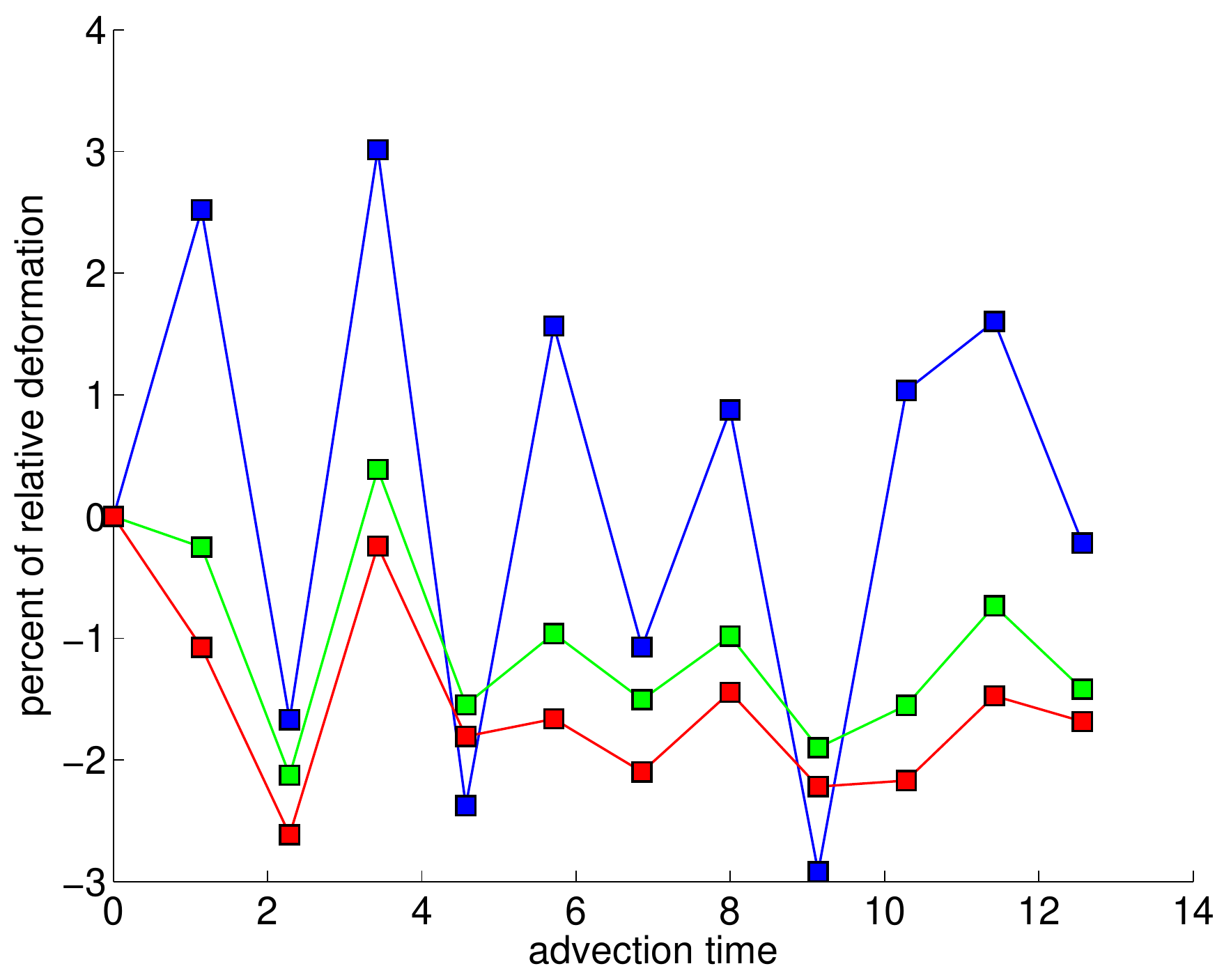} 
\caption{The relative stretching $\delta\ell(t)\times100$ of closed shearlines
of figure \ref{figure:shearlines-ex3}. The colors correspond to those
of figure \ref{figure:shearlines-ex3}. By their arc-length preservation
property, the advected elliptic barriers must theoretically have the
same arclength at times $t_{0}=0$ and $t_{1}=4\pi$. The numerical
error in arclength conservation is small overall, but more noticeable
for oscillations with large amplitudes (green and red curves of the right panel).}
\label{figure:length-ex3} 
\end{figure}

Figure~\ref{figure:shearlines-ex3} confirms that generalized KAM-type
curves, obtained as elliptic barriers, do exist in this problem. These
barriers are computed over the time interval $[0,4\pi]$ (i.e. $t_{0}=0$
and $t_{1}=t_{0}+T=4\pi$). As discussed in section \S\ref{sec:gtheory},
the arclength of an elliptic barrier at the initial time $t_{0}$
is equal to the arclength of its advected image under the flow map
$F_{t_{0}}^{t_{1}}$ at the final time $t_{1}$. This arclength preservation
is illustrated numerically in figure~\ref{figure:length-ex3}, which
shows the relative stretching,
\begin{equation}
\delta\ell(t)=\frac{\ell(\gamma_{t})-\ell(\gamma_{0})}{\ell(\gamma_{0})}
\label{eq:rel_def}
\end{equation}
of the time-$t$ image $\gamma_{t}$ of an elliptic barrier $\gamma_{0}$,
with $\ell$ referring to the arclength of the curve. Ideally,
the relative stretching of each elliptic barrier should be zero at time
$t_{1}=4\pi$, i.e. $\delta\ell(4\pi)=0$. Instead, we find that the relative
stretching $\delta\ell(4\pi)$ of the computed elliptic barriers is
at most $1.5\%$. This deviation from zero arises from numerical errors
in the computation of the CG strain tensor $C_{t_{0}}^{t_{1}}$ ,
which in turn causes small inaccuracies in the computation of closed
shearlines.

\begin{figure}[H]
\begin{centering}
\includegraphics[scale=0.5]{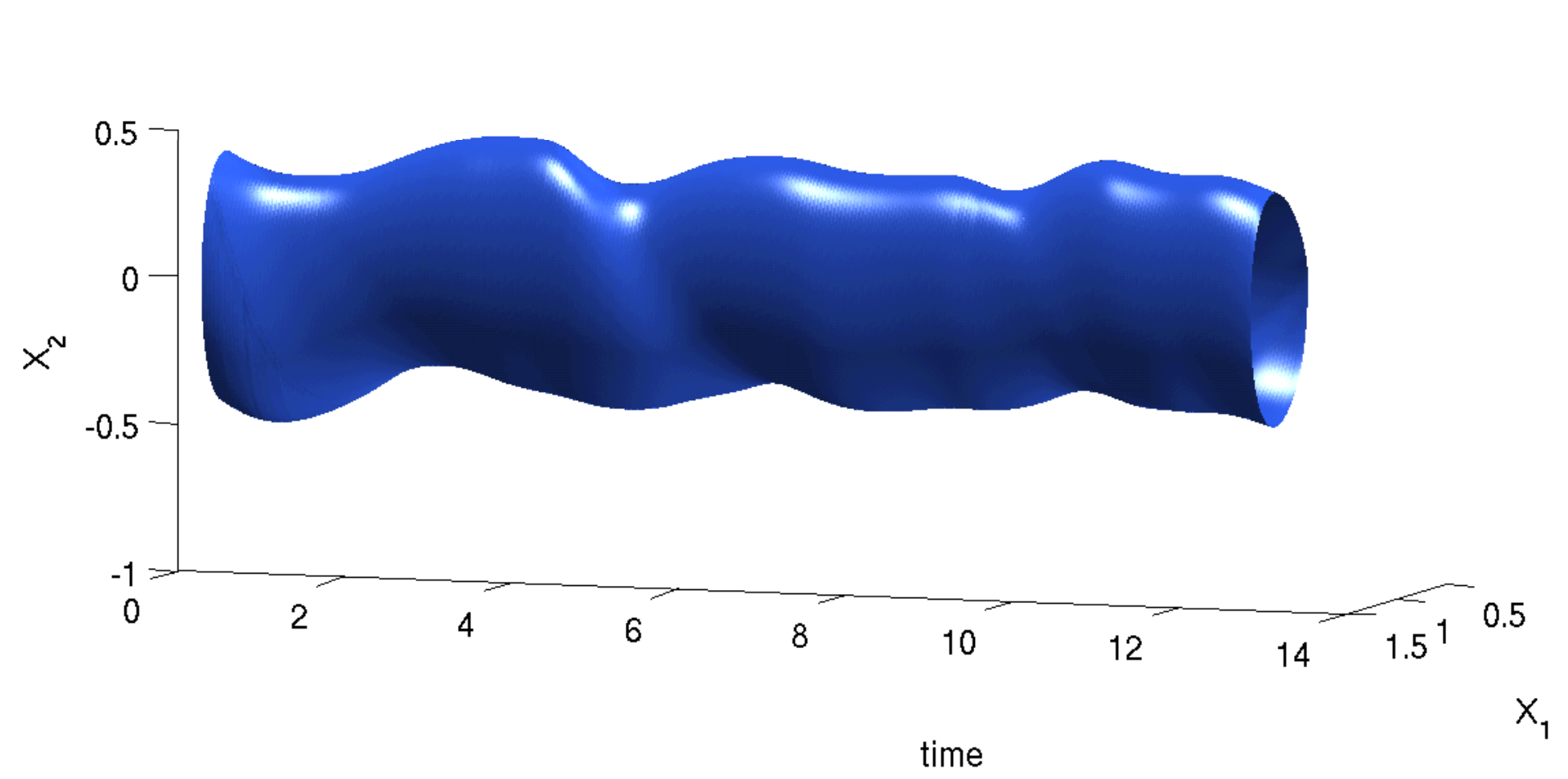} 
\par\end{centering}
\caption{Generalized KAM-type cylinder in the extended phase space of the aperiodically
forced Duffing undamped oscillator. The cylinder is obtained by advection of the closed shearline
shown in blue in figure~\ref{figure:shearlines-ex3}(b).}
\label{figure:area} 
\end{figure}

As noted earlier, the small relative stretching and the conservation
of enclosed area for an elliptic barrier in incompressible flow only
allows for small deformations when the barrier is advected in time.
This is illustrated in figure \ref{figure:area}, which shows the
blue elliptic barrier of figure \ref{figure:shearlines-ex3}b in the
extended phase-space. Each constant-time slice of the figure is the
advected image of the barrier.

Finally, we point out that the stability of the trajectories inside
elliptic barriers show a similar trend as in the case of the periodically
forced Duffing equation (figures \ref{figure:stability} and \ref{figure:stability-x1}).
Namely, perturbations inside the elliptic regions remain small
while they grow significantly inside the hyperbolic regions.

\subsubsection*{Case 2: Aperiodic forcing with damping }

In this final example, we consider the aperiodically forced, damped
Duffing oscillator 
\begin{eqnarray}
\dot{x}_{1} & = & x_{2},\nonumber \\
\dot{x}_{2} & = & x_{1}-x_{1}^{3}-\delta x_{2}+f(t),\label{eq:duffing-ex4}
\end{eqnarray}
with damping coefficient $\delta=0.15$. The forcing function
$f(t)$ is similar to that of Case I above, but with an amplitude twice
as large. As a result, none of the elliptic barriers survive even
in the absence of damping.

\begin{figure}[h]
\centering
\subfloat[]{\includegraphics[width=0.5\textwidth]{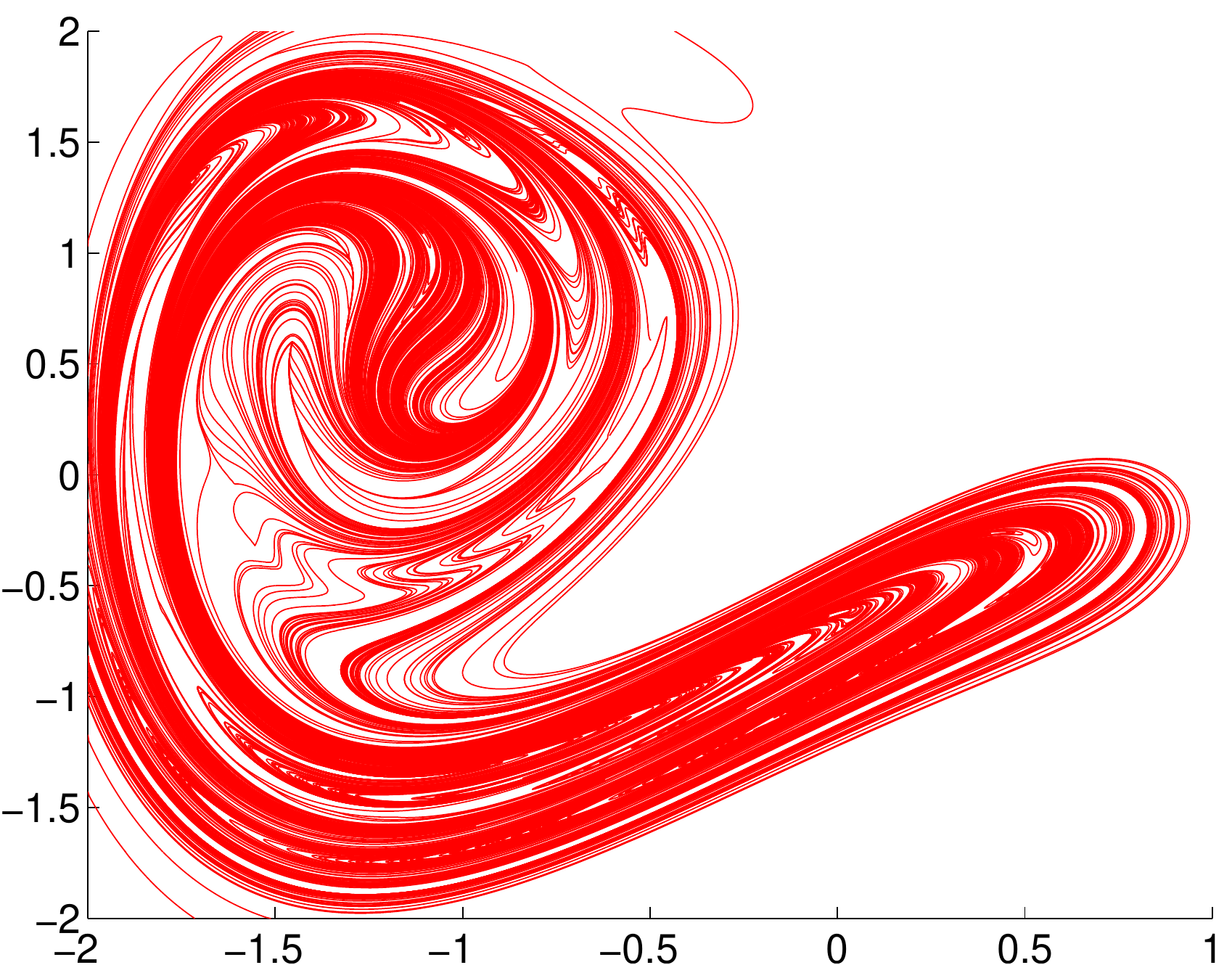}}
\subfloat[]{\includegraphics[width=0.5\textwidth]{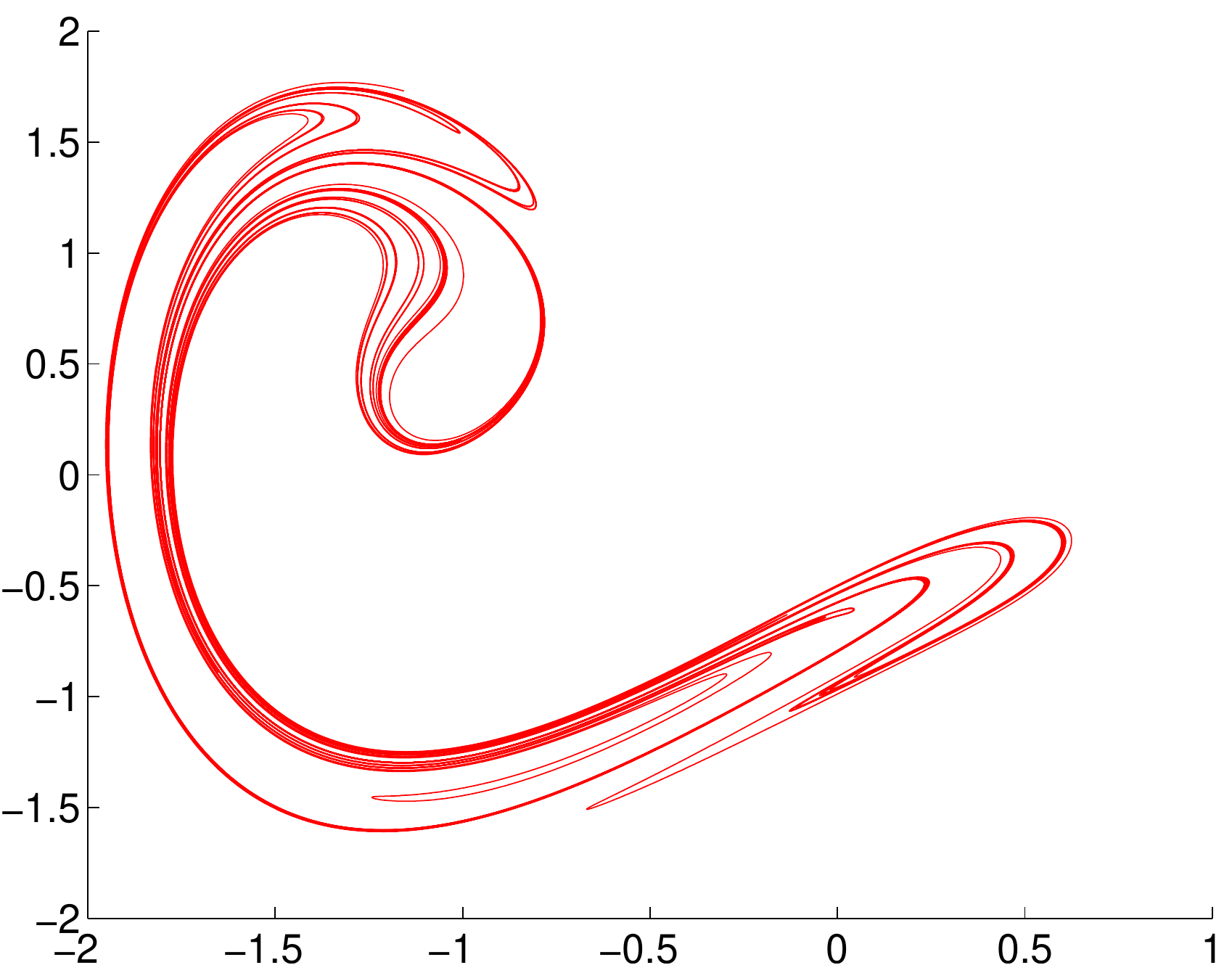}}
\caption{(a) Strainlines computed in backward time from $t_{0}=30$ to $t_{1}=10$.
(b) The resulting hyperbolic barrier extracted with maximum admissible
geodesic deviation of $\epsilon_{\xi_{1}}=10^{-5}$.}

\label{figure:strainlines-ex4} 
\end{figure}

Again, because of the aperiodic forcing, the behavior of this system
is a priori unknown and cannot be explored using Poincar\'{e} maps. In
order to investigate the existence of an attractor, strainlines (figure~\ref{figure:strainlines-ex4}a)
are computed from the backward-time CG strain tensor $C_{t_{0}}^{t_{1}}$
with $t_{0}=30$ and $t_{1}=10$. The strainline with minimum relative
stretching (\ref{eq:rel_stretch}) is then extracted. The part of
this strainline satisfying $d_{g}^{\xi_{1}}<\epsilon_{\xi_{1}}$ is
considered as the most influential hyperbolic barrier (figure~\ref{figure:strainlines-ex4}b).
The admissible upper bound $\epsilon_{\xi_{1}}$ for the geodesic
deviation is fixed as $10^{-5}$.

\begin{figure}[h]
\centering %
\begin{tabular}{cc}
 & \tabularnewline
\subfloat[]{\includegraphics[width=0.5\textwidth]{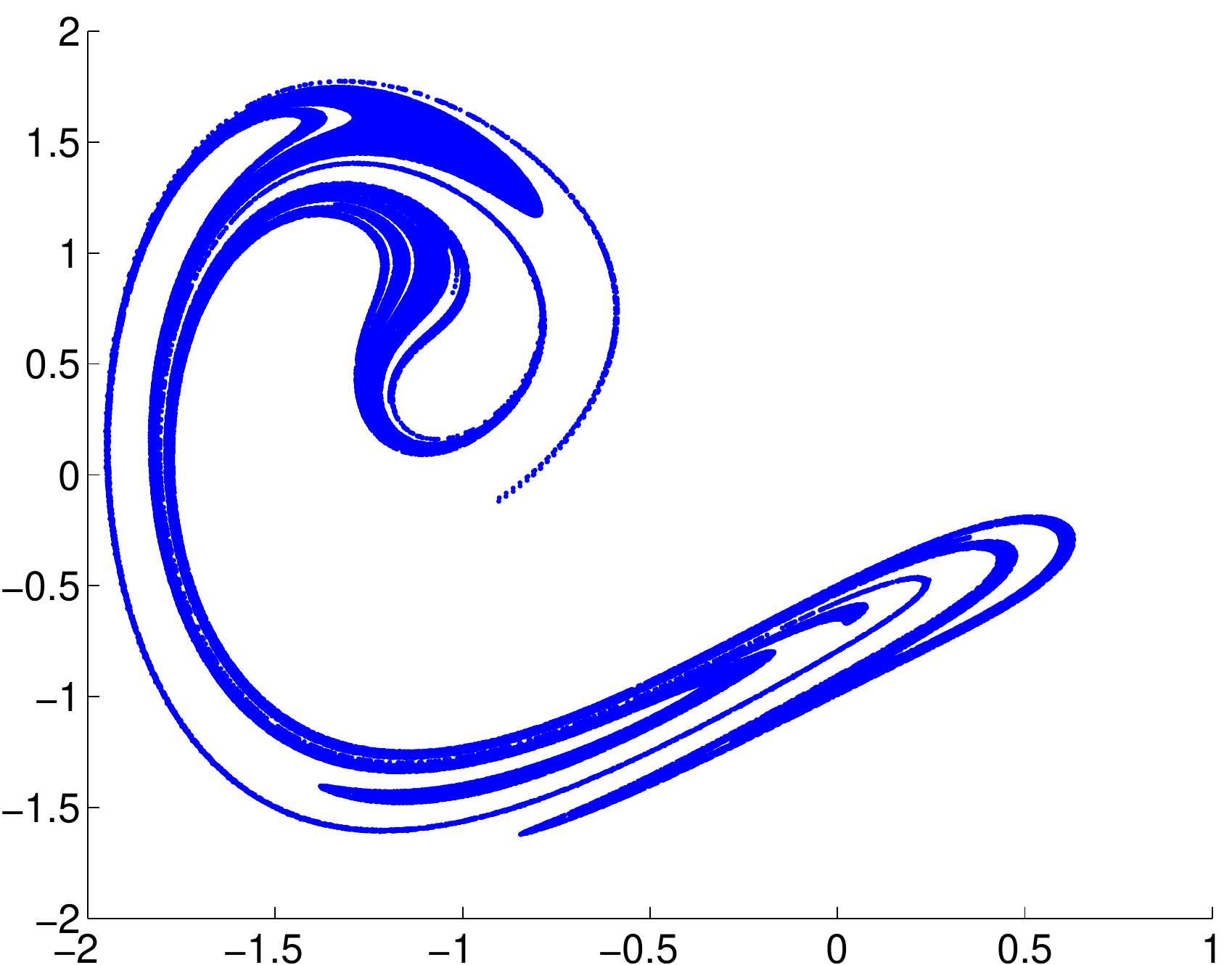}}
\subfloat[]{\includegraphics[width=0.5\textwidth]{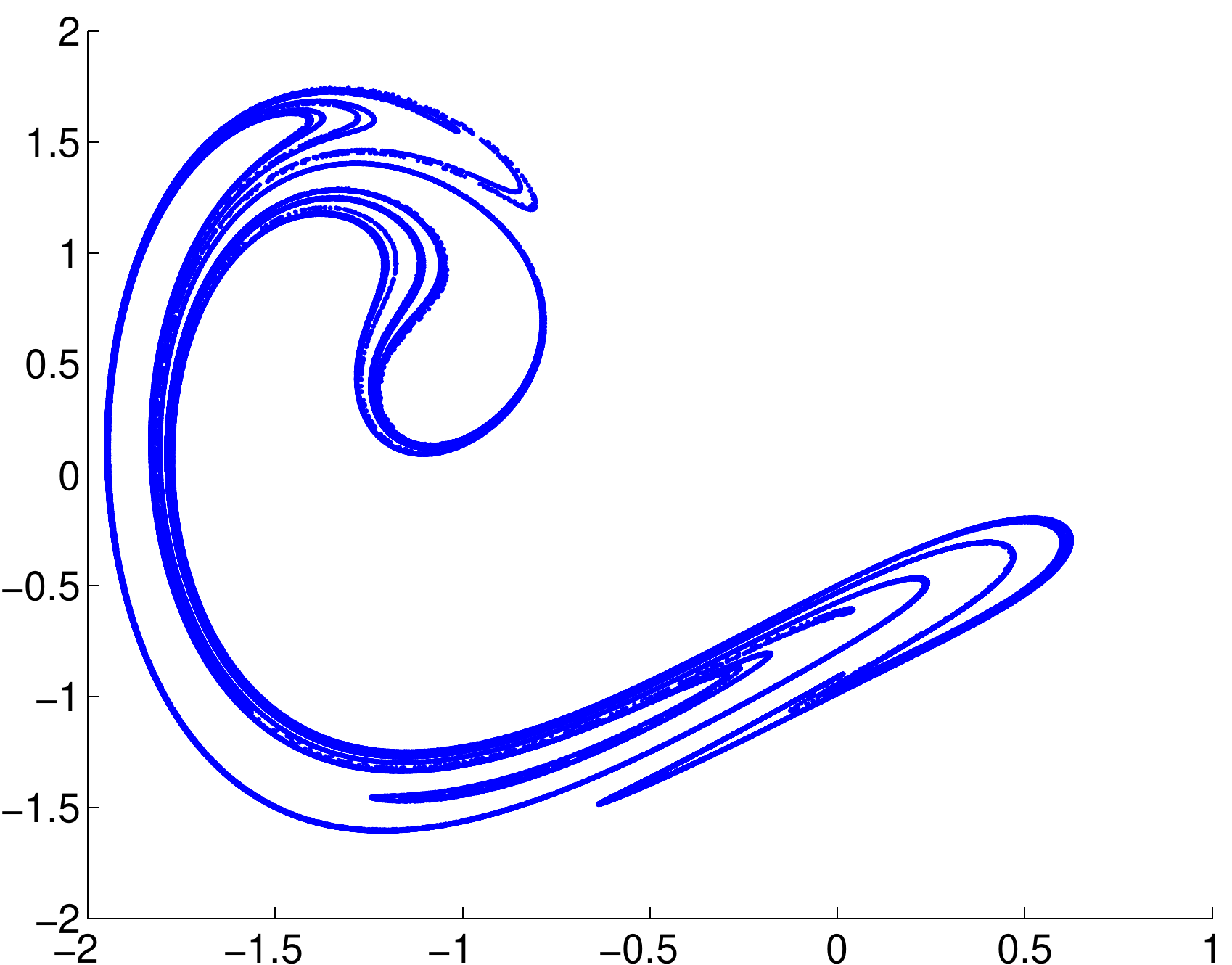}}
& \tabularnewline
\subfloat[]{\includegraphics[width=0.5\textwidth]{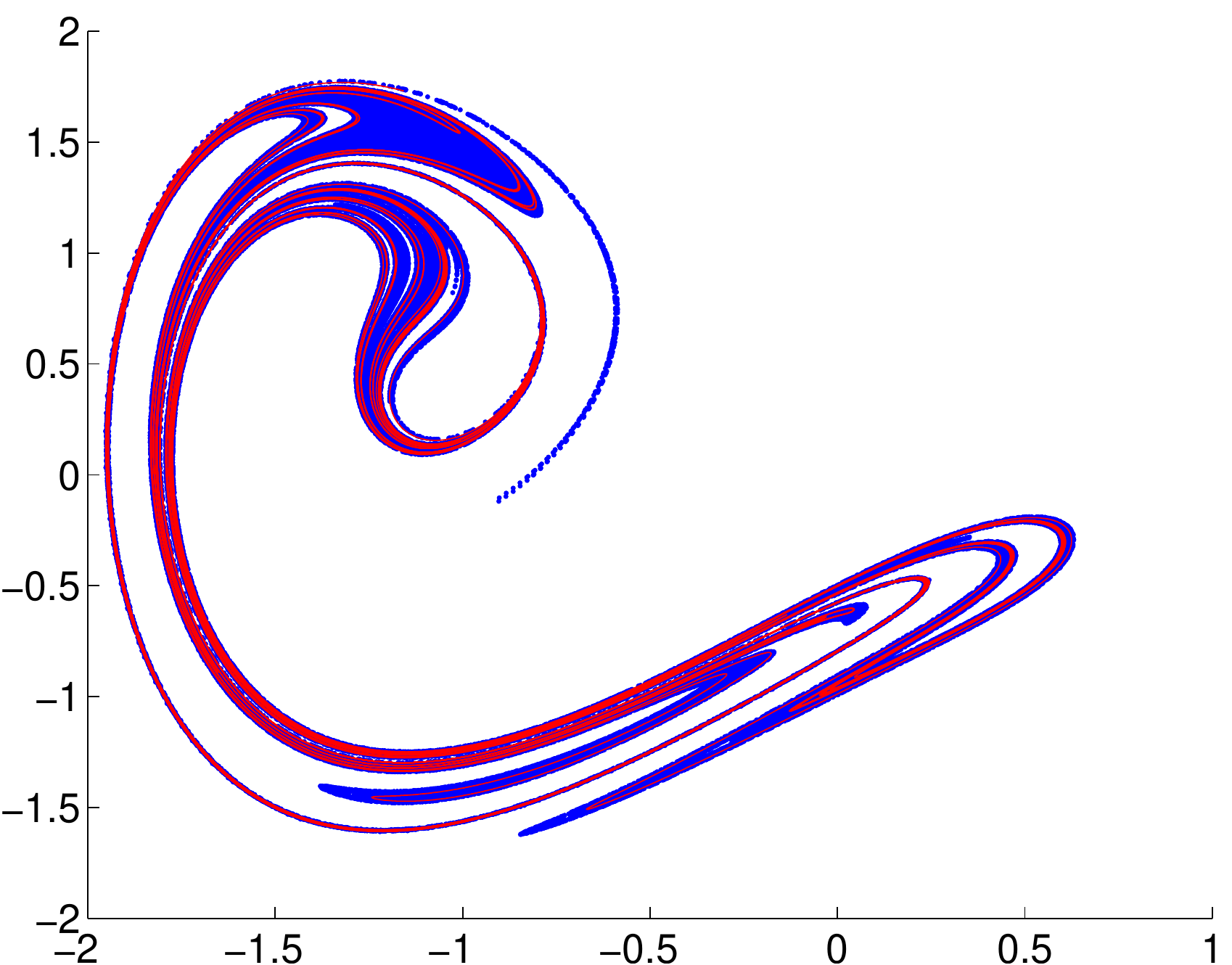}}

\subfloat[]{\includegraphics[width=0.5\textwidth]{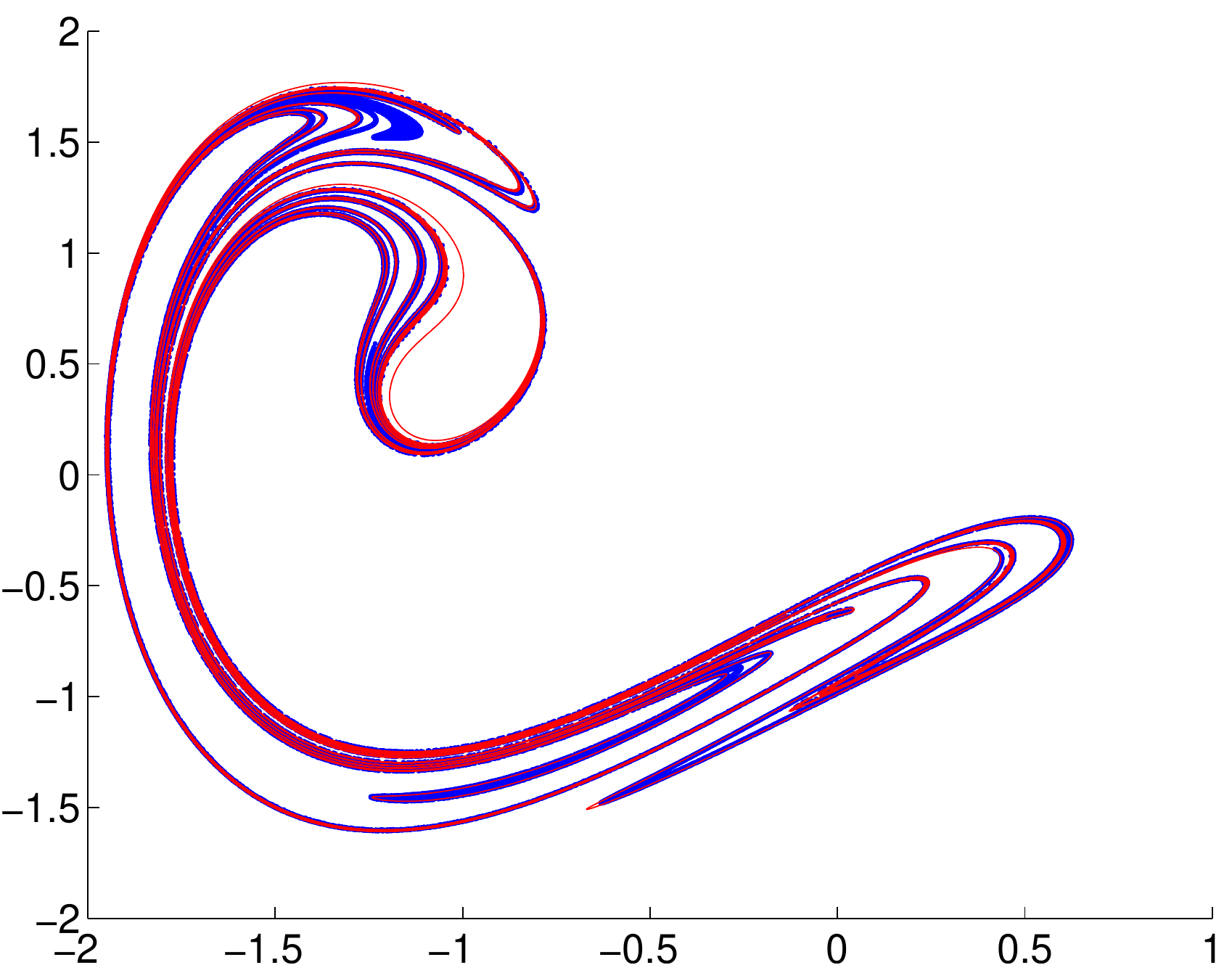}} & \tabularnewline
\end{tabular}\caption{(a) Tracers advected over the time interval from $t_1=10$ to $t_0=30$. (b) Tracers advected over a longer time interval from $t_1=0$ to $t_0=30$. (c)  The hyperbolic barrier (red) superimposed on the tracers advected
for the same time interval (d) Comparison of the hyperbolic barrier
(red) with the tracers advected for the longer time interval.}
\label{figure:compare-ex4} 
\end{figure}

In order to confirm the existence of the extracted attractor, we advect tracer particles
in forward time, first from time $t_{1}=10$ to time $t_{0}=30$,
then from $t_{1}=0$ to time $t_{0}=30$. Because of the fast-varying dynamics and weak dissipation,
a relatively long advection time is required for the tracers to converge
to the attractor. Figure \ref{figure:compare-ex4} shows the evolution
of tracers over $[t_{1},t_{0}].$ Note that the attractor inferred
from the tracers is less well pronounced than the hyperbolic barrier
extracted over the same length of time. This shows a clear advantage
for geodesic transport theory over simple numerical experiments with
tracer advection. For a longer integration time from $t_0=0$ to $t=30$, the tracers eventually converge to
the hyperbolic barrier.

\begin{figure}
\centering \includegraphics[width=0.6\textwidth]{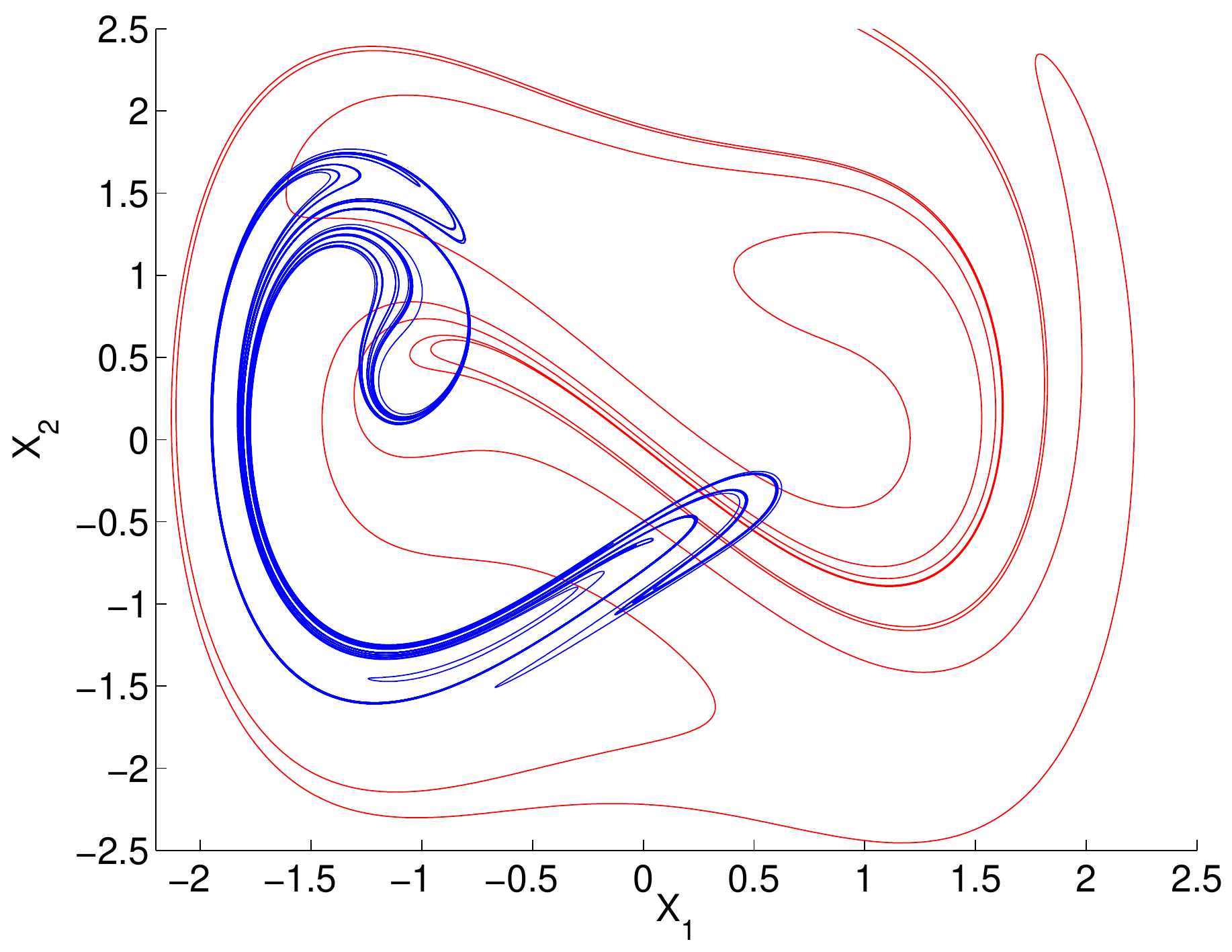}
\caption{Attracting (blue) and repelling (red) barriers at $t_{0}=30$ extracted
from backward-time and forward-time computations, respectively.}

\label{figure:lcs-ex4} 
\end{figure}

Repelling hyperbolic barriers can be computed similarly using forward-time
computations. Figure \ref{figure:lcs-ex4} shows both hyperbolic barriers
(stable and unstable manifolds) at time $t_{0}=30$. The repelling
barrier is computed from the CG strain tensor $\vc C_{t_{0}}^{t_{1}}$
with $t_{0}=30$ and $t_{1}=50$.

\section{Summary and Conclusions}

We have shown how the recently developed geodesic theory of transport
barriers \cite{haller12-1} in fluid flows can be adapted to compute 
finite-time invariant sets in one-degree-of-freedom mechanical
systems with general forcing. Specifically, in the presence of general
time dependence, temporally aperiodic stable- and unstable manifolds,
attractors, as well as generalized KAM tori can be located as hyperbolic
and elliptic barriers, respectively. The hyperbolic barriers are computed
as distinguished \emph{strainlines}, i.e. material lines along which the
Lagrangian strain is locally maximized. The elliptic barriers, on
the other hand, appear as distinguished \emph{shearlines}, i.e. material
lines along which the Lagrangian shear is locally maximized. The barriers
are finally identified as strainlines and shearlines that are most closely
approximated by least-stretching geodesics of the metric induced by
Cauchy--Green strain tensor.

We have used four simple examples for illustration. First, as benchmarks,
we considered periodically forced Duffing equations for which stable
and unstable manifolds, attractors and KAM curves can also be obtained
as invariant sets of an appropriately defined Poincar\'{e} map. We have
shown that elliptic barriers, computed as closed shearlines, coincide
with the KAM curves. Also, stable and unstable manifolds, as well
as attractors, can be recovered as hyperbolic barriers. More precisely,
as the integration time $T=t_1-t_0$ of the Cauchy--Green strain tensor $\vc C_{t_{0}}^{t_{1}}$
increases, the elliptic barriers in the periodically forced Duffing
equations converge to KAM curves. Similarly, the chaotic attractor
of the periodically forced and damped Duffing equation is more and
more closely delineated by a hyperbolic barrier computed from the
backward-time Cauchy--Green strain tensor $\vc C_{t_{1}}^{t_{0}}$
for increasing $T=t_0-t_1$ where $t_0>t_1$.

In the second set of examples, we have computed similar structures
for an aperiodically forced Duffing oscillator with and without damping.
In this case, Poincar\'{e} maps are no longer well-defined for the system,
and hence we had to advect tracer particles to verify the predictions
of the geodesic theory. Notably, tracer advection takes longer time
to reveal the structures in full detail than the geodesic theory does. Also,
tracer advection is only affective as a visualization tool if it relies
on a small number of particles, which in turn assumes that one already
roughly knows the location of the invariant set to be visualized.
Finally, unlike scattered tracer points, geodesic barriers are recovered
as parametrized smooth curves that provide a solid foundation for
further analysis or highly accurate advection.

In our examples, elliptic barriers have shown themselves as borders
of subsets of the phase-space that barely deform over time. In fact,
as illustrated in figure \ref{figure:stability}, outermost elliptic
barriers define the boundary between chaotic and regular dynamics.
Trajectories initiated inside elliptic barriers remain confined and
robust with respect to small perturbations. We believe that this property
could be exploited for stabilizing mechanical systems with general
time dependence. For instance, formulating an optimal control problem
for generating elliptic behavior in a desired part of the phase-space
is a possible approach.

Undoubtedly, the efficient and accurate computation of invariant
sets as geodesic transport barriers requires dedicated computational
resources. Smart algorithms reducing the computational cost are clearly
of interest. Parallel programming (both at CPU and GPU levels) has
previously been employed for Lagrangian coherent structure calculations
and should be useful in the present setting as well (see e.g. \cite{gpu-lcs}).
Other adaptive techniques are also available to lower the numerical
cost by reducing the computations to regions of interest (see e.g.
\cite{garth,lipinski10-1}).

In principle, invariant sets in higher-degree-of-freedom mechanical
systems could also be captured by similar techniques as locally least-stretching
surfaces. The development of the underlying multi-dimensional theory and computational
platform, however, is still underway. 

\section*{Acknowledgments}
M. F. would like to thank the Department of Mathematics at McGill University where this research was partially carried out. G. H. acknowledges partial support by the Canadian NSERC under grant 401839-11. 

\bibliographystyle{plain} 

\begin{thebibliography}{10}

\bibitem{Guckenheimer}  Guckenheimer, J., Holmes, P.: Nonlinear oscillations, dynamical systems, and bifurcations of vector fields. Springer-Verlag, (1990)

\bibitem{Strogatz} Strogatz, S.H.: Nonlinear Dynamics And Chaos. Westview Press, (2008) 

\bibitem{lcs_hurricane} Rutherford, B. and Dangelmayr, G., A three-dimensional Lagrangian hurricane eyewall computation. Q.J.R. Meteorol. Soc. 136, 1931–1944 (2010).

\bibitem{peacock10-1} Peacock, T., Dabiri, J.: Introduction to Focus Issue: Lagrangian Coherent Structures. Chaos 20(1) (2010)

\bibitem{Lai} Lai, Y.C., Tél, T.: Transient Chaos: Complex Dynamics on Finite Time Scales. Springer, (2011) 

\bibitem{haller12-1} Haller, G., Beron-Vera, F.J.: Geodesic theory of transport barriers in two-dimensional flows. Physica D: Nonlinear Phenomena 241(20), 1680-1702 (2012)

\bibitem{Truesdell} Truesdell, C., Noll, W., Antman, S.: The Non-Linear Field Theories of Mechanics. vol. v. 3. Springer, (2004) 

\bibitem{haller11-1} Haller, G.: A variational theory of hyperbolic Lagrangian Coherent Structures. Physica D-Nonlinear Phenomena 240(7), 574-598 (2011)

\bibitem{farazmand12-2} Farazmand, M., Haller, G.: Erratum and addendum to ``A variational theory of hyperbolic Lagrangian coherent structures (vol 240, pg 574, 2011).'' Physica D-Nonlinear Phenomena 241(4), 439-441 (2012)

\bibitem{farazmand12-1} Farazmand, M., Haller, G.: Computing Lagrangian coherent structures from their variational theory. Chaos 22(1) (2012) 

\bibitem{haller00-1} Haller, G., Yuan, G.: Lagrangian coherent structures and mixing in two-dimensional turbulence. Physica D: Nonlinear Phenomena 147(3–4), 352-370 (2000) 

\bibitem{faraz12_jfm} Farazmand, M., Haller, G., Geodesic transport barriers in two-dimensional turbulence, J.
Fluid Mech., 2012 (preprint) 

\bibitem{arnold78}  Arnold, V.I.: Mathematical Methods of Classical Mechanics. Springer, (1989) 

\bibitem{gpu-lcs} Garth, C., Li, C.S., Tricoche, X., Hansen, C.D., Hageni, H.: Visualization of Coherent Structures in Transient 2D Flows. In: Hege, H.C., Polthier, K., Scheuermann, G. (eds.) Topology-Based Methods in Visualization Ii. Mathematics and Visualization, pp. 1-13. (2009)

\bibitem{garth} Barakat, S., Garth, C., Tricoche, X.: Interactive Computation and Rendering of Finite-Time Lyapunov Exponent Fields. Ieee Transactions on Visualization and Computer Graphics 18(8), 1368-1380 (2012)

\bibitem{lipinski10-1} Lipinski, D., Cardwell, B., Mohseni, K.: A Lagrangian analysis of a two-dimensional airfoil with vortex shedding. Journal of Physics a-Mathematical and Theoretical 41(34) (2008)

\end{thebibliography}

\end{document}